%% file: GOF_FLM_MAR_arXiv_v2.tex
\documentclass[oneside,11pt]{article}

\input{preamble_arXiv.tex}

\usepackage[pdftex,final,bookmarksnumbered,bookmarksopen=false,breaklinks,colorlinks]{hyperref}
\hypersetup{
	final,
	bookmarksnumbered=true,
	bookmarksopen=true,
	bookmarksopenlevel=0,
	unicode=false,
	pdftoolbar=true,
	pdfmenubar=true,
	pdffitwindow=false,
	pdftitle={Testing for linearity in scalar-on-function regression with responses missing at random},
	pdfdisplaydoctitle=true,
	pdftoolbar=true,
	pdfmenubar=true,
	pdflang={English},
	pdfauthor={Manuel Febrero-Bande, Pedro Galeano, Eduardo Garc\'ia-Portugu\'es, Wenceslao Gonzalez-Manteiga},
	pdfsubject={arXiv paper},
	pdfcreator={Eduardo Garcia-Portugues},
	pdfproducer={Eduardo Garcia-Portugues},
	pdfkeywords={Functional linear model}{Functional principal components}{Goodness-of-fit tests}{Marked empirical processes}{Missing at random}{Wild bootstrap},
	pdfnewwindow=true,
	breaklinks=true,
	hidelinks,       
	linkcolor=black,
	citecolor=black,
	filecolor=magenta,
	urlcolor=cyan
}

\newif\ifmain
\maintrue
\newif\ifsupplement
\supplementtrue

\newif\iffigstabs
\figstabstrue

\begin{document}

\ifmain

\title{Testing for linearity in scalar-on-function regression \\ with responses missing at random}
\setlength{\droptitle}{-1cm}
\predate{}%
\postdate{}%
\date{}

\author{Manuel Febrero-Bande$^{1}$, Pedro Galeano$^{2}$, Eduardo Garc\'ia-Portugu\'es$^{2,3}$,\\ and Wenceslao Gonz\'alez-Manteiga$^{1}$}
\footnotetext[1]{Department of Statistics, Mathematical Analysis and Optimization, Universidade de Santiago de Compostela (Spain).}
\footnotetext[2]{Department of Statistics and UC3M-Santander Big Data Institute, Universidad Carlos III de Madrid (Spain).}
\footnotetext[3]{Corresponding author. e-mail: \href{mailto:edgarcia@est-econ.uc3m.es}{edgarcia@est-econ.uc3m.es}.}
\maketitle

\begin{abstract}
	A goodness-of-fit test for the Functional Linear Model with Scalar Response (FLMSR) with responses Missing at Random (MAR) is proposed in this paper. The test statistic relies on a marked empirical process indexed by the projected functional covariate and its distribution under the null hypothesis is calibrated using a wild bootstrap procedure. The computation and performance of the test rely on having an accurate estimator of the functional slope of the FLMSR when the sample has MAR responses. Three estimation methods based on the Functional Principal Components (FPCs) of the covariate are considered. First, the \textit{simplified} method estimates the functional slope by simply discarding observations with missing responses. Second, the \textit{imputed} method estimates the functional slope by imputing the missing responses using the simplified estimator. Third, the \textit{inverse probability weighted} method incorporates the missing response generation mechanism when imputing. Furthermore, both cross-validation and LASSO regression are used to select the FPCs used by each estimator. Several Monte Carlo experiments are conducted to analyze the behavior of the testing procedure in combination with the functional slope estimators. Results indicate that estimators performing missing-response imputation achieve the highest power. The testing procedure is applied to check for linear dependence between the average number of sunny days per year and the mean curve of daily temperatures at weather stations in Spain.
\end{abstract}

\begin{flushleft}
	\small\textbf{Keywords:} Functional linear model; Functional principal components; Goodness-of-fit tests; Marked empirical processes; Missing at random; Wild bootstrap.
\end{flushleft}

\section{Introduction}
\label{sec:Intro}

Functional Data Analysis (FDA) is a field of statistics that analyzes random variables that are observed throughout a continuum, typically in the form of random functions. For comprehensive introductions to the field, see \cite{Ramsay2005}, \cite{Ferraty2006}, \cite{Horvath2012}, \cite{Hsing2015}, and \cite{Kokoszka2017}. An important issue in FDA is the analysis of regression models where the response and/or covariates have a functional nature; see, for instance, the surveys by \cite{Febrero-Bande2017} and \cite{Reiss2017}, who focused on estimation methods when there is a functional covariate and a scalar response, and \cite{Greven2017}, who proposed a general framework for functional additive (mixed) models. The most analyzed model is the one given by $Y=m\left(\Xcal\right)+\varepsilon$, where $Y$ is a real response, $\Xcal$ is a functional covariate, $m\left(\xcal\right)=\mathbb{E}\left[Y|\Xcal=\xcal\right]$ is the regression function of $Y$ on $\Xcal$, and $\varepsilon$ is an error random variable with zero mean and finite variance $\sigma_{\varepsilon}^{2}$ that is uncorrelated with $\Xcal$.

When an independent and identically distributed (iid) random sample $\{\left(\Xcal_i, Y_i\right)\}_{i=1}^n$ from $\left(\Xcal, Y\right)$ is available, it is of interest to check whether the regression function $m$ is linear, conceptually being the simplest nontrivial parametric structure for $m$. For that, \cite{Garcia-Portugues:flm} proposed a goodness-of-fit test based on a Cram\'er--von Mises-like norm that integrates the marked empirical processes indexed by the projected functional covariate on a set of functional directions. As the resulting test is computationally intensive, \cite{Cuesta-Albertos:gofflm} considered the aforementioned empirical process but for a collection of \textit{randomly-chosen} directions; goodness-of-fit tests then follow by applying norms over the projected processes, with a posterior aggregation of the resulting $p$-values. The projection-based procedure is computationally less expensive than the Cram\'er--von Mises-like procedure, although at the cost of some loss of power. From a different perspective, \cite{McLean2015} considered testing for linearity within the Functional Generalized Additive Model (FGAM), a nonlinear extension of the Functional Linear Model with Scalar Response (FLMSR). To this end, these authors formulate the FGAM as a mixed model using a representation for penalized splines with three pairwise independent vectors of random effects. Then, the problem is reduced to one of testing for zero variance components in a mixed model and a restricted likelihood ratio statistic is used to conduct the test. When the response is functional, \cite{Chen2020} (for scalar predictors) and \cite{Garcia-Portugues2020} (for functional predictor) considered goodness-of-fit tests based on a Cram\'er--von Mises-like norm. The latter paper uses a double-projected empirical process and wild bootstrap for calibration, whereas the former uses a nonparametric Monte Carlo method for calibration. When the covariates are also scalar and the response is functional, still within the functional linear model, \cite{Smaga2022} considered tests about general linear hypotheses based on random projections. For more information on goodness-of-fit advances for functional regression models, not only for linear specifications, see the reviews in \cite{Gonzalez-Manteiga:reviewfunc} and \cite{Gonzalez-Manteiga2023}.

Sometimes, the observed random sample contains responses that are Missing At Random (MAR). For instance, \cite{Febrero-Bande2019} analyzed the mean curves of the annual average daily temperature to predict the corresponding average number of sunny days per year at several weather stations in Spain. The responses contained missing values due to the unavailability of recordings for the presence of opaque clouds at certain weather stations, so these authors proposed two methods to estimate the functional slope of the FLMSR. The first one is the \textit{simplified} method: it estimates the functional slope using only the observed covariate-response pairs and then uses this estimation to impute missing responses and predict new responses. The \textit{imputed} method re-estimates the functional slope using both complete pairs and pairs with responses imputed by the simplified method to subsequently impute and predict. \cite{Crambes2018} obtained mean square error rates for the imputed values resulting from the simplified functional slope estimator. When the regression function operator is unspecified, \cite{Ferraty2013} and \cite{Ling2016} focused on estimating the unconditional mean and the conditional mode of the response, respectively, when some of the responses are MAR. \cite{Ling2015} developed a nonparametric method to estimate the regression function operator under MAR responses and \cite{Ling2022} investigated the estimation of the functional single index regression model with MAR responses for strong mixing time series data. \cite{ZhuZhao2023} proposed diagnostic measures to identify influential observations in the FLMSR when observations on the scalar response are subject to nonignorable missingness. From an application standpoint, \cite{Ciarleglio2022} related certain brain measures derived from electroencephalography through functional regression and proposed several imputation methods for the missing measure observations. Beyond FDA, the MAR case has been thoroughly studied in multiple linear regression; see \cite{Gonzalez-Manteiga2006}, \cite{Sun2009}, \cite{Li2012}, \cite{Sun2017}, \cite{Zheng2020}, \cite{PerezGonzalez2021}, and references therein.

The main contribution of this paper is a testing procedure to check whether the scalar-on-function regression function $m$ is linear when the sample contains MAR responses. The procedure follows the guidelines set out in \cite{Garcia-Portugues:flm}, rather than those in \cite{Cuesta-Albertos:gofflm}, because, although less computationally efficient, the former testing procedure is empirically more powerful than the latter. Furthermore, as a dataset with MAR responses has some limited information, it seems convenient to gain power at the cost of losing computational efficiency. The proposed testing procedure relies on a statistic constructed from a marked empirical process based on residuals obtained after estimating the functional slope with a MAR-adapted estimator. To this end, three estimation methods based on the Functional Principal Components (FPCs) of the covariate are used to estimate the functional slope of the FLMSR when the sample includes MAR responses. Their performances are compared in terms of mean square error of estimation and computational cost through several Monte Carlo experiments. In addition to the simplified and imputed estimators considered in \cite{Febrero-Bande2019}, the \textit{inverse probability weighted} estimator is introduced. Like the imputed method, it imputes missing responses before estimating the functional slope, but, differently, it takes into account the mechanism of missing response generation to perform the imputation. Both Cross-Validation (CV) and LASSO regression are used to select the FPCs of the covariate used by each estimator. Thus, six estimators are actually considered, resulting from the combination of the three estimation methods and the two methods for selecting the FPCs. Several Monte Carlo experiments suggest that the highest power of the testing procedure is achieved with estimators that perform imputation of missing responses; simply discarding pairs of observations with missing responses is suboptimal in terms of power. The testing procedure shows that there is no evidence to reject the FLMSR in the real-data application considered in \cite{Febrero-Bande2019}.

The rest of this paper is structured as follows. Section~\ref{sec:TestingProblem} introduces the testing problem and the main assumptions when the scalar response is MAR. Section~\ref{sec:Estimation-FLM-MAR} presents estimators of the functional slope of the FLMSR with MAR responses suitable for performing the linearity test. Section~\ref{sec:TestingProcedure} provides the testing procedure and describes the wild bootstrap resampling strategy. Section~\ref{sec:MonteCarlos} illustrates the performance of the testing procedure in several Monte Carlo experiments. Section~\ref{sec:Real} presents a real-data application of the test. Finally, Section~\ref{sec:Conclusions} summarizes some concluding remarks.

\section{The testing problem}
\label{sec:TestingProblem}

Let $\mathbb{H}$ be a real separable Hilbert space endowed with the inner product $\langle\cdot,\cdot\rangle$ and its associated norm $\|\cdot\|$. The most common setup in the literature is that $\mathbb{H}$ is the $L^2\left[a,b\right]$ space defined as the set of all functions $\eta:\left[a,b\right]\rightarrow\mathbb{R}$ such that $\|\eta\|:=\langle\eta,\eta\rangle^{1/2}$ is finite, where the inner product of two functions $\eta,\varphi\in L^2\left[a,b\right]$ is defined as $\langle\eta,\varphi\rangle:=\int_{a}^{b}\eta\left(t\right)\varphi\left(t\right)\,\mathrm{d}t$. Furthermore, let $\Xcal$ be a functional random variable in $\mathbb{H}$ with zero-mean function and covariance operator $\Gamma$ defined as
\begin{align}\label{CovOp}
	\Gamma\left[\eta \right]:=\mathbb{E}\left[\left(\Xcal\otimes\Xcal\right)\left(\eta\right)\right]:=\mathbb{E}\left[\left\langle\Xcal,\eta\right\rangle\Xcal\right],\quad \text{for any $\eta \in \mathbb{H}$,}
\end{align}
that is assumed to have a sequence of positive eigenvalues $a_1>a_2>\cdots>0$ associated with a set of orthonormal eigenfunctions $\psi_{1},\psi_{2},\ldots $ such that $\Gamma\left[\psi_{k}\right]=a_{k}\psi_{k}$, for $k\in\mathbb{N}$. 

It is assumed that the functional random variable $\Xcal$ is related to a zero-mean real random variable $Y$, defined in the same probability space, through the regression model $Y=m\left(\Xcal\right)+\varepsilon$, where $m:\xcal\in\mathbb{H}\mapsto\mathbb{E}\left[Y|\Xcal=\xcal\right]\in\mathbb{R}$ is the regression function of $Y$ over $\Xcal$ and $\varepsilon$ is a zero-mean random variable with finite variance $\sigma_{\varepsilon}^{2}$ that is uncorrelated with $\Xcal$. Furthermore, $Y$ has some potential missingness, as later described in \eqref{MAR}. Within this setting, the goal is to test the composite hypothesis
\begin{align}  \label{Null-Hypothesis}
	H_0:m\in \mathcal{L}:=\left\{\langle\cdot,\beta\rangle:\beta\in \mathbb{H}\right\}
\end{align}
versus the general alternative hypothesis
\begin{align}  \label{Alternative-Hypothesis}
	H_{1}:m\not\in \mathcal{L}:=\left\{\langle\cdot,\beta\rangle:\beta\in \mathbb{H}\right\}.
\end{align}
Thus, if the null hypothesis \eqref{Null-Hypothesis} holds, the relationship between the functional covariate $\Xcal$ and the real response $Y$ is driven by the FLMSR given by
\begin{align}  \label{FLM-1}
	Y=\left\langle \Xcal,\beta \right\rangle+\varepsilon,
\end{align}
where $\beta\in \mathbb{H}$ is the unspecified functional slope of the model, whose existence and uniqueness are ensured, as shown in \cite{Cardot2007}, under the following two fundamental assumptions:
\begin{enumerate}
	\item[A1.] $\Xcal$ and $Y$ satisfy $\sum\limits_{k=1}^{\infty}\frac{1}{a_{k}^{2}}\left\langle \mathbb{E}\left[Y\Xcal\right],\psi_{k}\right\rangle^{2}<\infty$;
	\item[A2.] $\mathrm{Ker}\left(\Gamma\right)=\left\{0\right\}$.
\end{enumerate}

Now, to test the null hypothesis \eqref{Null-Hypothesis} versus the alternative hypothesis \eqref{Alternative-Hypothesis}, it is assumed that an iid random sample $\left\{ \left( \Xcal_{i},Y_{i},R_{i}\right)\right\}_{i=1}^n$ generated from the triplet $\left( \Xcal,Y,R\right) $ is available, where $R$ is a Bernoulli variable such that $R_{i}=1$ if $Y_{i}$ is observed and $R_{i}=0$ if $Y_{i}$ is missing, for $i=1,\ldots,n$. More precisely, the missing mechanism is assumed to be MAR, with the observance probability driven by
\begin{align}\label{MAR}
	\mathbb{P}\left(R=1|Y,\Xcal\right)=\mathbb{P}\left(R=1|\Xcal\right)=:p\left(\Xcal\right),
\end{align}
where $p:\mathbb{H}\rightarrow[0,1]$ is an unspecified function operator of $\Xcal$ that describes the missing probability for $Y$. Hence, the real response $Y$ and the indicator variable $R$ are conditionally independent given the functional covariate $\Xcal$, a common and realistic mechanism that allows missing responses to be predicted with the available information. If \eqref{MAR} does not hold, then the values of the missing responses cannot be adequately predicted as they depend on information that is not available.

\section{Parameter estimation of the FLMSR with MAR responses}
\label{sec:Estimation-FLM-MAR}

The MAR-adapted procedure presented in Section~\ref{sec:TestingProcedure} for testing \eqref{Null-Hypothesis} relies crucially on an appropriate estimator of the functional slope $\beta$ of the FLMSR \eqref{FLM-1} and the residuals associated with that estimator. Hence, the simplified and imputed estimators proposed in \cite{Febrero-Bande2019} are briefly summarized next and then the inverse probability weighted estimator is proposed, which, unlike the previous ones, takes into account the mechanism of generation of missing responses in \eqref{MAR}. Furthermore, following \cite{Garcia-Portugues2020}, the simplified, imputed, and inverse probability weighted \textit{MAR-adapted LASSO-selected} estimators are presented, which use LASSO to estimate the FLMSR instead of Ordinary Least Squares and Cross-Validation (OLS+CV), as done in \cite{Febrero-Bande2019}. Thus, an array of six estimators is available, which will be compared in Section~\ref{sec:MonteCarlos}.

For all this, it is helpful to write the FLMSR \eqref{FLM-1} in a simpler way. Under the assumptions given in Section \ref{sec:TestingProblem}, the functional random variable $\Xcal$ and the functional slope $\beta$ can be represented in terms of the eigenfunctions of the covariance operator $\Gamma$ in \eqref{CovOp} as follows:
\begin{align}\label{Chi-Beta}
	\Xcal =\sum\limits_{k=1}^{\infty}S_{k}\psi_{k},\quad \beta=\sum\limits_{k=1}^{\infty}b_{k}\psi_{k},
\end{align}
where $S_{k}:=\left\langle \Xcal ,\psi_{k}\right\rangle$, $k\in\mathbb{N}$, are the FPC scores, which are uncorrelated random variables with $\mathbb{E}\left[S_{k}\right]=0$ and $\mathbb{V}\mathrm{ar}\left[S_{k}\right]=a_{k}$, and $b_{k}:=\left\langle \beta ,\psi_{k}\right\rangle$, $k\in\mathbb{N}$, are deterministic coefficients. Therefore, the FLMSR \eqref{FLM-1} can be written as
\begin{align}\label{FLM-2}
	Y=\sum_{k=1}^{\infty}b_{k}S_{k}+\varepsilon,
\end{align}
which allows $\beta$ to be estimated in three steps (see \cite{Cardot2007} when the sample is fully observed and \cite{Febrero-Bande2019} when the sample contains MAR responses): (\textit{i}) \textit{truncate} the series \eqref{Chi-Beta} and \eqref{FLM-2} up to $K_{n}\leq n$ terms, where $K_{n}$ may depend on the sample; (\textit{ii}) \textit{plug-in} estimators $\big\{\widehat{\psi}_k\big\}_{k=1}^{K_{n}}$ of $\big\{\psi_k\big\}_{k=1}^{K_{n}}$ in $S_k$; and (\textit{iii}) \textit{fit} the resultant linear model. More precisely, whether or not the responses are missing, the covariance operator $\Gamma$ in \eqref{CovOp} is estimated with the sample covariance operator of $\left\{\Xcal_{i}\right\}_{i=1}^{n}$ defined as
\begin{align}\label{CovOp-Sample}
	\widehat{\Gamma}\left[\eta \right]:=\frac{1}{n}\sum\limits_{i=1}^{n}\left\langle \Xcal_{i},\eta \right\rangle \Xcal_{i},\quad \text{for any }\eta \in \mathbb{H},
\end{align}
that has a sequence of nonnegative eigenvalues $\widehat{a}_{1}\geq \widehat{a}_{2}\geq \cdots \geq 0$, with $\widehat{a}_{k}=0$, for $k>n$, and a set of orthonormal eigenfunctions $\widehat{\psi}_{1},\widehat{\psi}_{2},\ldots $ such that $\widehat{\Gamma}\big[\widehat{\psi}_{k}\big]=\widehat{a}_{k}\widehat{\psi}_{k}$, for $k=1,2,\ldots$ Similarly, the sample FPC scores are defined as $\widehat{S}_{i,k}:=\big\langle\Xcal_{i},\widehat{\psi}_{k}\big\rangle$, for $i=1,\ldots,n$ and $k=1,\ldots,K_{n}$, giving rise to the linear model associated with~\eqref{FLM-2}:
\begin{align}\label{FLM-3}
	Y_{i}=\sum\limits_{k=1}^{K_{n}}b_{k}\widehat{S}_{i,k}+\varepsilon_{i},\quad \text{for $i=1,\ldots,n$.}
\end{align}

When there are MAR responses, the \textit{simplified} estimator follows the naive idea of estimating $\{b_k\}_{k=1}^{K_{n}}$ in \eqref{FLM-3} using only the pairs of observed responses and FPC scores. More precisely, the simplified estimator of $\beta $ is given by
\begin{align}\label{Beta-S}
	\widehat{\beta}_{\mathcal{S}}:=\sum\limits_{k=1}^{K_{\mathcal{S}}}\widehat{b}_{k,\mathcal{S}}\widehat{\psi}_{k},
\end{align}
where $\widehat{b}_{1,\mathcal{S}},\ldots,\widehat{b}_{K_{\mathcal{S}},\mathcal{S}}$ are the OLS estimators of $b_1,\ldots,b_{K_{\mathcal{S}}}$, i.e.,
\begin{align*}
	\widehat{b}_{k,\mathcal{S}}:=\frac{1}{n_{\mathcal{S}}\widehat{a}_{k}}\sum\limits_{i\in i_{\mathcal{S}}}Y_{i}\widehat{S}_{i,k},\quad\text{for $k=1,\ldots,K_{\mathcal{S}}$},
\end{align*}
with $i_{\mathcal{S}}:=\left\{i:R_{i}=1\right\}_{i=1}^{n}$, $n_{\mathcal{S}}:=\#\left\{i_{\mathcal{S}}\right\}$, $K_{\mathcal{S}}$ a certain cutoff selected from the set $\left\{1,\ldots,K_{\max}\right\}$ by leave-one-out CV, and $K_{\max}\leq n$ a certain upper bound. 

The \textit{imputed} estimator imputes missing responses with the simplified estimator $\widehat{\beta}_{\mathcal{S}}$ in \eqref{Beta-S} to then estimate $\{b_k\}_{k=1}^{K_{n}}$ in \eqref{FLM-3} with all pairs of responses (observed or imputed) and FPC scores. The equality $\mathbb{E}\left[RY+\left(1-R\right)\left\langle \Xcal,\beta\right\rangle\right]=\mathbb{E}\left[Y\right]$, similar to the one used in \cite{Sun2009} and \cite{Sun2017}, among others, in multiple linear regression models, justifies the use of the completed sample $\left\{\left(\Xcal_{i},Y_{i,\mathcal{S}},R_{i}\right)\right\}_{i=1}^n$ to estimate $\beta$, where $Y_{i,\mathcal{S}}:=R_{i}Y_{i}+\left(1-R_{i}\right)\big\langle \Xcal_{i},\widehat{\beta}_{\mathcal{S}}\big\rangle$, for $i=1,\ldots,n$. Then, the imputed estimator of $\beta $ is defined as
\begin{align}\label{Beta-I}\widehat{\beta}_{\mathcal{I}}:=\sum\limits_{k=1}^{K_{\mathcal{I}}}\widehat{b}_{k,\mathcal{I}}\widehat{\psi}_{k},
\end{align}
where $K_{\mathcal{I}}$ is a cutoff and $\widehat{b}_{1,\mathcal{I}},\ldots,\widehat{b}_{K_{\mathcal{I}},\mathcal{I}}$ are the OLS estimators of $b_1,\ldots,b_{K_{\mathcal{I}}}$, i.e.,
\begin{align*}
	\widehat{b}_{k,\mathcal{I}}:=\frac{1}{n\widehat{a}_{k}}\sum\limits_{i=1}^{n}Y_{i,\mathcal{S}}\widehat{S}_{i,k},\quad\text{for $k=1,\ldots,K_{\mathcal{I}}$},
\end{align*}
where the cutoffs $\left(K_{\mathcal{S}},K_{\mathcal{I}}\right)$ are selected from the set $\left\{1,\ldots,K_{\max}\right\}^2$ by leave-one-out CV. Importantly, the cutoff $K_{\mathcal{S}}$ chosen for $\widehat{\beta}_{\mathcal{S}}$ does not have to be the same as the cutoff $K_{\mathcal{S}}$ chosen for $\widehat{\beta}_{\mathcal{I}}$. \cite{Febrero-Bande2019} evidence that $\widehat{\beta}_{\mathcal{I}}$ can have a lower mean square error of estimation than $\widehat{\beta}_{\mathcal{S}}$ if the cutoffs $\left(K_{\mathcal{S}},K_{\mathcal{I}}\right)$ are selected jointly. 

The \textit{inverse probability weighted} estimator is based on the equality $\mathbb{E}\big[\left(R/p\left(\Xcal\right)\right)Y+\break\left(1-R/p\left(\Xcal\right)\right)\big\langle \Xcal,\beta\big\rangle\big]=\mathbb{E}\left[Y\right]$, where $p\left(\cdot\right)$ is the missing response operator \eqref{MAR}, which is similar to the one used in \cite{Sun2009}, \cite{Sun2017}, \cite{Qin2017}, and \cite{Bianco2019,Bianco2020}, among others, in multiple linear regression models. This justifies the completed sample $\left\{\left(\Xcal_{i},Y_{i,\mathcal{W}},R_{i}\right)\right\}_{i=1}^n$, where $Y_{i,\mathcal{W}}:=\left(R_i/\widehat{p}\left(\Xcal_i\right)\right)Y_i+\left(1-R_i/\widehat{p}\left(\Xcal_i\right)\right)\big\langle \Xcal_{i},\widehat{\beta}_{\mathcal{S}}\big\rangle$ for $i=1,\ldots,n$, and $\widehat{p}\left(\cdot\right)$ is the local constant Nadaraya--Watson estimator of the MAR mechanism $p\left(\cdot\right)$,
\begin{align}\label{Estimator-NW}
	\widehat{p}\left(\xcal\right):=\sum_{i=1}^{n} \frac{K_h\left(\left\|\xcal-\Xcal_i\right\|\right)}{\sum_{l=1}^{n}K_h\left(\left\|\xcal-\Xcal_l\right\|\right)}R_i.
\end{align}
Above, $K_h\left(\cdot\right):=K\left(\cdot/h\right)/h$, $K:\left[0,\infty\right)\to\left[0,\infty\right)$ is a kernel and $h$ is a bandwidth that can be chosen by CV. The inverse probability weighted estimator of $\beta$ is defined~as
\begin{align}\label{Beta-W}\widehat{\beta}_{\mathcal{W}}:=\sum\limits_{k=1}^{K_{\mathcal{W}}}\widehat{b}_{k,\mathcal{W}}\widehat{\psi}_{k},
\end{align}
where $K_{\mathcal{W}}$ is a cutoff and $\widehat{b}_{1,\mathcal{W}},\ldots,\widehat{b}_{K_{\mathcal{W}},\mathcal{W}}$ are the OLS estimators of $b_1,\ldots,b_{K_{\mathcal{W}}}$, i.e.,
\begin{align*}
	\widehat{b}_{k,\mathcal{W}}:=\frac{1}{n\widehat{a}_{k}}\sum\limits_{i=1}^{n}Y_{i,\mathcal{W}}\widehat{S}_{i,k},\quad\text{for $k=1,\ldots,K_{\mathcal{W}}$},
\end{align*}
where the cutoffs $\left(K_{\mathcal{S}},K_{\mathcal{W}}\right)$ are chosen similarly to the imputed estimator cutoffs.

When all covariates and responses are observed, \cite{Garcia-Portugues2020} proposed a LASSO-selected estimator of $\beta$ to estimate the coefficients $b_k$ and select significant FPC scores in \eqref{FLM-3}. This estimator is particularly relevant when $\beta$ projects into a large number of FPCs, thus producing a high-dimensional regression problem. Therefore, to have a larger array of competitive estimators, the simplified, imputed, and inverse probability weighted MAR-adapted LASSO-selected estimators of $\beta$ are introduced next. First, LASSO-type regression in $\big\{\big(Y_i,\widehat{S}_{i,k}\big):i\in i_{\mathcal{S}}\big\}$ is carried out to obtain
\begin{align*}
	\big(\widetilde{b}_{1,\mathcal{SL}},\ldots,\widetilde{b}_{K_{\max},\mathcal{SL}}\big)^{\prime}:=\underset{(b_{1},\ldots,b_{K_{\max}})'\in \mathbb{R}^{K_{\max}}}{\arg\min}\left\{\sum\limits_{i\in i_{\mathcal{S}}}\left( Y_{i}-\sum\limits_{k=1}^{K_{\max}}b_{k}\widehat{S}_{i,k}\right)^{2}+\lambda\sum_{k=1}^{K_{\max}}\left|b_k\right|\right\}
\end{align*}
where $\lambda\geq 0$ is the penalty parameter that is efficiently selected through CV and the so-called one standard error rule \cite[see, e.g.,][]{Friedman2010}. Then, the coefficients $b_k$ selected by LASSO are re-estimated using OLS to reduce the mean square error of the estimation, leading to the \textit{simplified MAR-adapted LASSO-selected} estimator of $\beta$:
\begin{align}\label{Beta-SL}
	\widehat{\beta}_{\mathcal{SL}}:=\sum\limits_{k\in \mathcal{K}_{\mathcal{SL}}}\widehat{b}_{k,\mathcal{SL}}\widehat{\psi}_{k},
\end{align}
where $\mathcal{K}_{\mathcal{SL}}$ is the subset of values of $k\in\left\{1,\ldots,K_{\max}\right\}$ such that $\widetilde{b}_{k,\mathcal{SL}}\neq 0$, and
\begin{align}\label{betas-SL}
	\widehat{b}_{k,\mathcal{SL}}:=\frac{1}{n_{\mathcal{S}}\widehat{a}_{k}}\sum\limits_{i\in i_{\mathcal{S}}}Y_{i}\widehat{S}_{i,k},\quad\text{for $k\in \mathcal{K}_{\mathcal{SL}}$},
\end{align}
are the OLS of $\left\{b_k:k\in \mathcal{K}_{\mathcal{SL}}\right\}$. Second, the completed samples $\left\{\left(\Xcal_{i}, Y_{i,\mathcal{IL}}, R_{i}\right)\right\}_{i=1}^n$ and $\left\{\left(\Xcal_{i}, Y_{i,\mathcal{WL}}, R_{i}\right)\right\}_{i=1}^n$, where $Y_{i,\mathcal{IL}}:=R_{i}Y_{i}+\left(1-R_{i}\right)\big\langle \Xcal_{i},\widehat{\beta}_{\mathcal{SL}}\big\rangle$ and $Y_{i,\mathcal{WL}}:=\left(R_i/\widehat{p}\left(\Xcal_i\right)\right)Y_i+\left(1-R_i/\widehat{p}\left(\Xcal_i\right)\right)\big\langle \Xcal_{i},\widehat{\beta}_{\mathcal{SL}}\big\rangle$, lead to the \textit{imputed} and the \textit{inverse probability weighted MAR-adapted LASSO-selected} estimators, respectively given by
\begin{align}\label{Beta-IL}
	\widehat{\beta}_{\mathcal{IL}}:=\sum\limits_{k\in \mathcal{K}_{\mathcal{IL}}}\widehat{b}_{k,\mathcal{IL}}\widehat{\psi}_{k}
\end{align}
and
\begin{align}\label{Beta-WL}
	\widehat{\beta}_{\mathcal{WL}}:=\sum\limits_{k\in \mathcal{K}_{\mathcal{WL}}}\widehat{b}_{k,\mathcal{WL}}\widehat{\psi}_{k}.
\end{align}
Above, $\mathcal{K}_{\mathcal{IL}}$ and $\mathcal{K}_{\mathcal{WL}}$ are the equivalents of $\mathcal{K}_{\mathcal{SL}}$ in the simplified MAR-adapted LASSO-selected estimator. In the same way, $\widehat{b}_{k,\mathcal{IL}}$ and $\widehat{b}_{k,\mathcal{WL}}$ are the equivalents of $\widehat{b}_{k,\mathcal{SL}}$ in \eqref{betas-SL}.

Two comments on the proposed estimators of $\beta$ are in order. First, in LASSO regression it is common to standardize the covariates to avoid different scales. However, the FPC scores are not standardized here because their scaling comes from the process of constructing the FPCs. Second, the value of $K_{\max}$ is fixed as that which accumulates some amount of the predictor variability. In this way, FPCs that explain very little variability are ignored, avoiding a substantial increase in the mean square error of the estimation of $\beta$. More details are given in Section~\ref{sec:MonteCarlos}.

\section{The MAR-adapted testing procedure}
\label{sec:TestingProcedure}

The MAR-adapted procedure for testing the null hypothesis \eqref{Null-Hypothesis} versus the alternative hypothesis \eqref{Alternative-Hypothesis} with the sample $\left\{ \left( \Xcal_{i}, Y_{i}, R_{i}\right)\right\}_{i=1}^n$ is presented next. The procedure is based on a Cram\'er--von Mises statistic and follows the approach in \cite{Garcia-Portugues:flm}, which in turn builds on the proposal of \cite{Escanciano2006}, when all responses have been fully observed. For this, the characterizations of the null hypothesis \eqref{Null-Hypothesis} in terms of random projections of the functional covariate $\Xcal$ in the following lemma will be useful in the ongoing analysis.

\begin{lemma}[\citealp{Garcia-Portugues:flm}]\label{lem}
	Let $\beta$ be an element of $\mathbb{H}$, $\mathbb{S}_{\mathbb{H}}:=\left\{\gamma\in\mathbb{H}:\left\|\gamma\right\|=1\right\}$ and $\mathbb{S}_{\mathbb{H}}^{\mathcal{K}}:=\left\{\gamma=\sum_{k\in \mathcal{K}}\left\langle\gamma,\psi_{k}\right\rangle\psi_{k}\in\mathbb{H}:\left\|\gamma\right\|=1\right\}$, where $\mathcal{K}$ is any finite set of positive integers. Then, the following statements are equivalent:
	
	\begin{enumerate}[label=(\roman{*}), ref=(\emph{\roman{*}})]
		
		\item $H_{0}$ holds, that is, $m\left(\Xcal\right)=\left\langle\Xcal,\beta\right\rangle$, $\forall\Xcal\in\mathbb{H}$. \label{lem:1}
		
		\item $\mathbb{E}\left[\left(Y-\left\langle\Xcal,\beta\right\rangle\right)1_{\left\{\langle\Xcal,\gamma\rangle\leq u\right\}}\right]=0$, for a.e. $u\in \mathbb{R}$ and $\forall\gamma\in \mathbb{S}_{\mathbb{H}}$. \label{lem:2}
		
		\item $\mathbb{E}\left[\left(Y-\left\langle\Xcal,\beta\right\rangle\right)1_{\left\{\langle\Xcal,\gamma\rangle\leq u\right\}}\right]=0$, for a.e. $u\in \mathbb{R}$ and $\forall\gamma\in \mathbb{S}_{\mathbb{H}}^{\mathcal{K}}$. \label{lem:3}
		
	\end{enumerate}
	
\end{lemma}

As the null hypothesis \eqref{Null-Hypothesis} is characterized by the null value of $\mathbb{E}\left[\left(Y-\left\langle\Xcal,\beta\right\rangle\right)1_{\left\{\langle\Xcal,\gamma\rangle\leq u\right\}}\right]$, for a.e. $u\in \mathbb{R}$ and $\forall\gamma\in \mathbb{S}_{\mathbb{H}}$, the goal is to measure the deviation of the sample with MAR responses, $\left\{ \left( \Xcal_{i}, Y_{i}, R_{i}\right)\right\}_{i=1}^{n}$, from the null hypothesis \eqref{Null-Hypothesis} with such characterization. For this, a Residual Marked empirical Process based on Projections (RMPP) is defined with the residuals 
\begin{align}\label{Residuals}
	\widehat{\varepsilon}_{i}&=Y_{i}-\big\langle\Xcal_i,\widehat{\beta}\big\rangle,\quad i\in i_{\mathcal{S}},
\end{align}
where $\widehat{\beta}$ stands for any of the considered estimators of $\beta$, i.e., $\widehat{\beta}_{\mathcal{S}}$ in \eqref{Beta-S}, $\widehat{\beta}_{\mathcal{I}}$ in \eqref{Beta-I}, $\widehat{\beta}_{\mathcal{W}}$ in \eqref{Beta-W}, $\widehat{\beta}_{\mathcal{SL}}$ in \eqref{Beta-SL}, $\widehat{\beta}_{\mathcal{IL}}$ in \eqref{Beta-IL}, or $\widehat{\beta}_{\mathcal{WL}}$ in \eqref{Beta-WL}. Importantly, although imputation under the null hypothesis is performed to obtain $\widehat{\beta}_{\mathcal{I}}$, $\widehat{\beta}_{\mathcal{W}}$, $\widehat{\beta}_{\mathcal{IL}}$, and $\widehat{\beta}_{\mathcal{WL}}$, the residuals corresponding to the imputed responses are \textit{not} included in \eqref{Residuals}, thus preventing imputations under the null hypothesis from significantly affecting the results.

Now, the deviation of the random sample with MAR responses from the null hypothesis \eqref{Null-Hypothesis} is measured with a norm of the RMPP
\begin{align}\label{RMPPs-1}
	R_{n_{\mathcal{S}}}\big(\widehat{\beta},u,\gamma\big)=n_{\mathcal{S}}^{-1/2}\sum\limits_{i\in i_{\mathcal{S}}}\widehat{\varepsilon}_{i}1_{\left\{\left\langle\Xcal_{i},\gamma\right\rangle\leq u\right\}},
\end{align}
where $u\in \mathbb{R}$ and $\gamma\in \mathbb{S}_{\mathbb{H}}$, that is quantified through the Cram\'er--von Mises statistic
\begin{align}\label{CVMs-1}
	\mathrm{PCvM}_{n_{\mathcal{S}}}\big(\widehat{\beta}\big)=\int_{\mathbb{R}\times \mathbb{S}_{\mathbb{H}}}R_{n_{\mathcal{S}}}\big(\widehat{\beta},u,\gamma\big)^2F_{n_{\mathcal{S}},\gamma}\left(\mathrm{d}u\right)\omega\left(\mathrm{d}\gamma\right),
\end{align}
where $F_{n_{\mathcal{S}},\gamma}$ is the Empirical Cumulative Distribution Function (ECDF) of the set of projected covariates $\left\{\left\langle\Xcal_i,\gamma\right\rangle\right\}_{i\in i_{\mathcal{S}}}$, and $\omega$ represents a measure on $\mathbb{S}_{\mathbb{H}}$. Since the computation of \eqref{CVMs-1} is not practically feasible because $\gamma$ has an infinite dimension, two actions are taken. First, \ref{lem:3} in Lemma \ref{lem} is used to replace $\gamma\in \mathbb{S}_{\mathbb{H}}$ in \eqref{CVMs-1} with $\gamma\in \mathbb{S}_{\mathbb{H}}^{\mathcal{K}}$, for $\mathcal{K}$ being the set of FPCs used to construct $\widehat{\beta}$. Second, the RMPP \eqref{RMPPs-1} is evaluated at $\widehat{\gamma}=\sum_{k\in \mathcal{K}}\widehat{\gamma}_{k}\widehat{\psi}_{k}$, where $\widehat{\gamma}_{k}=\big\langle\widehat{\gamma},\widehat{\psi}_{k}\big\rangle$ for $k\in \mathcal{K}$, leading to the RMPP given by
\begin{align*}
	R_{n_{\mathcal{S}}}\big(\widehat{\beta},u,\widehat{\gamma}\big)=
	n_{\mathcal{S}}^{-1/2}\sum\limits_{i\in i_{\mathcal{S}}}\widehat{\varepsilon}_{i}1_{\left\{\left\langle\Xcal_{i},\widehat{\gamma}\right\rangle\leq u\right\}},
\end{align*}
that allows the modified statistic to be defined as
\begin{align*}
	\mathrm{PCvM}_{n_{\mathcal{S}},\mathcal{K}}\big(\widehat{\beta}\big)=\int_{\mathbb{R}\times \mathbb{S}_{\mathbb{H}}^{\mathcal{K}}}R_{n_{\mathcal{S}}}\big(\widehat{\beta},u,\widehat{\gamma}\big)^2F_{n_{\mathcal{S}},\widehat{\gamma}}\left(\mathrm{d}u\right)\omega\left(\mathrm{d}\widehat{\gamma}\right),
\end{align*}
where $F_{n_{\mathcal{S}},\widehat{\gamma}}$ is the ECDF of the set of projected covariates given by 
$\left\langle \Xcal_{i},\widehat{\gamma}\right\rangle=\left\langle\sum_{k=1}^{\infty}\widehat{S}_{i,k}\widehat{\psi}_{k},\sum_{k\in \mathcal{K}}\widehat{\gamma}_{k}\widehat{\psi}_{k}\right\rangle=\sum_{k\in \mathcal{K}}\widehat{S}_{i,k}\widehat{\gamma}_{k}$, for $i=1,\ldots,n$. Moreover, as in \cite{Garcia-Portugues:flm}, it is possible to show that
\begin{align}\label{CVMs-2}\mathrm{PCvM}_{n_{\mathcal{S}},\mathcal{K}}\big(\widehat{\beta}\big)=n_{\mathcal{S}}^{-2}\widehat{\varepsilon}^{\prime}A\widehat{\varepsilon},
\end{align}
where $\widehat{\varepsilon}=\left(\widehat{\varepsilon}_{\left[1\right]},\ldots,\widehat{\varepsilon}_{\left[n_{\mathcal{S}}\right]}\right)^{\prime}$ is the $n_{\mathcal{S}}$-dimensional vector containing the set of residuals $\left\{\widehat{\varepsilon}_{i}\right\}_{i\in i_{\mathcal{S}}}$ in \eqref{Residuals}, such that $\widehat{\varepsilon}_{\left[l\right]}$ is the $l$-th residual in $\left\{\widehat{\varepsilon}_{i}\right\}_{i\in i_{\mathcal{S}}}$, for $l=1,\ldots,n_{\mathcal{S}}$, and $A$ is a $n_{\mathcal{S}}\times n_{\mathcal{S}}$ matrix with entries $A_{lm}=\left(\sum_{r=1}^{n_{\mathcal{S}}}A_{lmr}\right)_{lm}$, where $A_{lmr}$, for $l,m,r=1,\ldots,n_{\mathcal{S}}$, is another matrix given by $A_{lmr}=\frac{\pi^{N_{\mathcal{K}}/2-1}}{\Gamma\left(N_{\mathcal{K}}/2\right)}A_{lmr}^{\left(0\right)}$, with $N_{\mathcal{K}}=\#\mathcal{K}$,
\begin{equation*}
	A_{lmr}^{\left( 0\right) }=
	\begin{cases}
		2\pi & \!\text{if }\widehat{S}_{\left[ l\right] ,\mathcal{K}}=\widehat{S}_{\left[ m\right] ,\mathcal{K}}=\widehat{S}_{\left[ r\right] ,\mathcal{K}}, \\
		\pi  &
		\!\begin{array}{l}
			\!\text{if }\widehat{S}_{\left[ l\right] ,\mathcal{K}}=\widehat{S}_{\left[ m\right] ,\mathcal{K}},\,\widehat{S}_{\left[ l\right] ,\mathcal{K}}=\widehat{S}_{\left[ r\right] ,\mathcal{K}}, \\
			\!\text{or }\widehat{S}_{\left[ m\right] ,\mathcal{K}}=\widehat{S}_{\left[ r\right] ,\mathcal{K}},
		\end{array}
		\\
		\left\vert \pi -\arccos \left( \frac{\left( \widehat{S}_{\left[ l\right] ,\mathcal{K}}-\widehat{S}_{\left[ r\right] ,\mathcal{K}}\right) ^{\prime
			}\left( \widehat{S}_{\left[ m\right] ,\mathcal{K}}-\widehat{S}_{\left[ r\right] ,\mathcal{K}}\right) }{\left\Vert \widehat{S}_{\left[ l\right] ,\mathcal{K}}-\widehat{S}_{\left[ r\right] ,\mathcal{K}}\right\Vert
			\left\Vert \widehat{S}_{\left[ m\right] ,\mathcal{K}}-\widehat{S}_{\left[ r\right] ,\mathcal{K}}\right\Vert }\right) \right\vert  & \!\text{else},
	\end{cases}
\end{equation*}
and $\widehat{S}_{\left[l\right],\mathcal{K}}$, for $l=1,\ldots,n_{\mathcal{S}}$, is the $N_{\mathcal{K}}$-dimensional column vector containing the sample functional component scores $\big\{\widehat{S}_{\left[l\right],k}\big\}_{k\in \mathcal{K}}$. Importantly, \eqref{CVMs-2} is a proper norm for the residuals \citep[Lemma 4]{Garcia-Portugues2020}.

A wild bootstrap procedure is used to calibrate the distribution of the statistic \eqref{CVMs-2} under the null hypothesis \eqref{Null-Hypothesis}. The whole testing procedure is described below.

\begin{algo}[\textbf{Cram\'er--von Mises testing procedure}] {\ }
	
	Let $\left\{\left(\Xcal_i,Y_i,R_i\right)\right\}_{i=1}^n$ be an iid sample from $\left(\Xcal,Y,R\right)$, where $Y=m\left(\Xcal\right)+\varepsilon$ and $R|\Xcal=\xcal\sim\mathrm{Ber}\left(p\left(\xcal\right)\right)$ is the MAR operator. The procedure to test $H_0:m\in\mathcal{L}:=\{\langle\cdot,\beta \rangle:\beta\in\mathbb{H}\}$ against $H_1:m\not\in\mathcal{L}$ is as follows:
	
	\begin{enumerate}
		
		\item Obtain the eigenfunctions $\big\{\widehat{\psi}_{k}\big\}_{k=1}^{K_{\max}}$ of the sample covariance operator $\widehat{\Gamma}$ in \eqref{CovOp-Sample}, for an upper bound $K_{\max}$.
		
		\item Compute the estimator $\widehat{\beta}$, which results in the set of FPC indices~$\mathcal{K}$.
		
		\item Obtain the residuals $\big\{\widehat{\varepsilon}_{i}\big\}_{i\in i_{\mathcal{S}}}$ in \eqref{Residuals} associated with $\widehat{\beta}$.
		
		\item Compute the statistic $\mathrm{PCvM}_{n_{\mathcal{S}},\mathcal{K}}\big(\widehat{\beta}\big)$ in \eqref{CVMs-2}.
		
		\item Bootstrap resampling. For $b=1,\ldots,B$, do:
		
		\begin{enumerate}
			
			\item Draw binary iid random variables $\big\{V_{i}^{*b}\big\}_{i\in i_{\mathcal{S}}}$ with $\mathbb{P}\big[V^{*b}=\big(1\mp\sqrt{5}\big)/2\big]=\big(5\pm\sqrt{5}\big)/10$.
			
			\item Construct bootstrap residuals $\widehat{\varepsilon}_{i}^{*b}=V_{i}^{*b}\widehat{\varepsilon}_{i}$ for $i\in i_{\mathcal{S}}$.
			
			\item Define a bootstrap sample $\big\{\big(\Xcal_i,Y_{i}^{*b},R_i\big)\big\}_{i=1}^{n}$ where $Y_{i}^{*b}=\big\langle\Xcal_i,\widehat{\beta}\big\rangle+\widehat{\varepsilon}_{i}^{*b}$ for $i\in i_{\mathcal{S}}$.
			
			\item Obtain the estimator $\widehat{\beta}^{*b}$ with the bootstrap sample $\big\{\big(\Xcal_i,Y_{i}^{*b},R_i\big)\big\}_{i=1}^{n}$.
			
			\item Obtain the set of residuals $\big\{\widehat{\varepsilon}_{i}^{*b}\big\}_{i\in i_{\mathcal{S}}}$ associated with $\widehat{\beta}^{*b}$.
			
			\item Compute the statistic $\mathrm{PCvM}_{n_{\mathcal{S}},\mathcal{K}}\big(\widehat{\beta}^{*b}\big)$ in \eqref{CVMs-2}.
			
		\end{enumerate}
		
		\item Estimate the $p$-value of the test with $\#\big\{\mathrm{PCvM}_{n,\mathcal{K}}\big(\widehat{\beta}\big)\leq \mathrm{PCvM}_{n,\mathcal{K}}\big(\widehat{\beta}^{\ast b}\big)\big\}/B$.
		
	\end{enumerate}
	
\end{algo}

Two comments about the testing procedure are in order. First, the symmetric matrix $A$ and the missing probability estimators $\widehat{p}\left(\Xcal_1\right),\ldots,\widehat{p}\left(\Xcal_n\right)$ need to be computed only once, drastically reducing the computational cost of the procedure. Second, the choice of $V^{*b}$ is the golden section bootstrap suggested by \cite{Mammen1993}, yet any other random variable $V^{*b}$ verifying $\mathbb{E}\big[V^{*b}\big]=0$ and $\mathbb{V}\mathrm{ar}\big[V^{*b}\big]=1$ can be used.

\section{Monte Carlo experiments}
\label{sec:MonteCarlos}

In this section, two Monte Carlo experiments are presented: (\textit{i}) to compare the finite sample properties of the estimators of $\beta$ presented in Section~\ref{sec:Estimation-FLM-MAR}; and (\textit{ii}) to explore the empirical performance of the proposed procedure to test the composite hypothesis \eqref{Null-Hypothesis} in terms of these estimators.

\subsection{Finite sample properties of the estimators of \texorpdfstring{$\beta$}{beta}}
\label{subsec:Estimators}

The first Monte Carlo experiment compares the finite sample properties of the simplified \eqref{Beta-S}, imputed \eqref{Beta-I}, and inverse probability weighted \eqref{Beta-W} estimators, and their respective LASSO-selected counterparts \eqref{Beta-SL}, \eqref{Beta-IL}, and \eqref{Beta-WL}. For this purpose, as in \cite{Garcia-Portugues:flm}, the FLMSR \eqref{FLM-1} is considered with $\beta$ being one of $\beta_1\left(t\right)=\sin\left(2\pi t\right)-\cos\left(2\pi t\right)$, $\beta_2\left(t\right)=t-\left(t-0.75\right)^2$, or $\beta_3\left(t\right)=t+\cos\left(2\pi t\right)$, for $t\in\left[0,1\right]$ (see Figure~\ref{fig:Betas-Model-1}), the functional covariate $\Xcal$ is an Ornstein--Uhlenbeck process in $\left[0,1\right]$ with $\mathbb{E}\left[\Xcal\left(t\right)\right]=0$ and $\mathbb{C}\mathrm{ov}\left[\Xcal\left(s\right),\Xcal\left(t\right)\right]=\frac{3}{2}\left(\exp\left(\frac{2}{3}\min\left\{s,t\right\}\right)-1\right)$, for $s,t\in\left[0,1\right]$, and $\varepsilon$ is Gaussian with mean $0$ and standard deviation $\sigma_{\varepsilon}=0.1$. With these characteristics, the resulting coefficients of determination, as defined in \cite{Febrero-Bande2017}, are those that appear in the first column in Table~\ref{tab:Coefficients_Determination}. Both functional slopes and covariates, as well as any other functional object in this section, are represented in the form of $201$ equidistant points in the interval $\left[0,1\right]$. 

\begin{figure}[!htbp]
	\centering
	\includegraphics[width=.66\textwidth,clip,trim={0cm 0.5cm 0cm 1.25cm}]{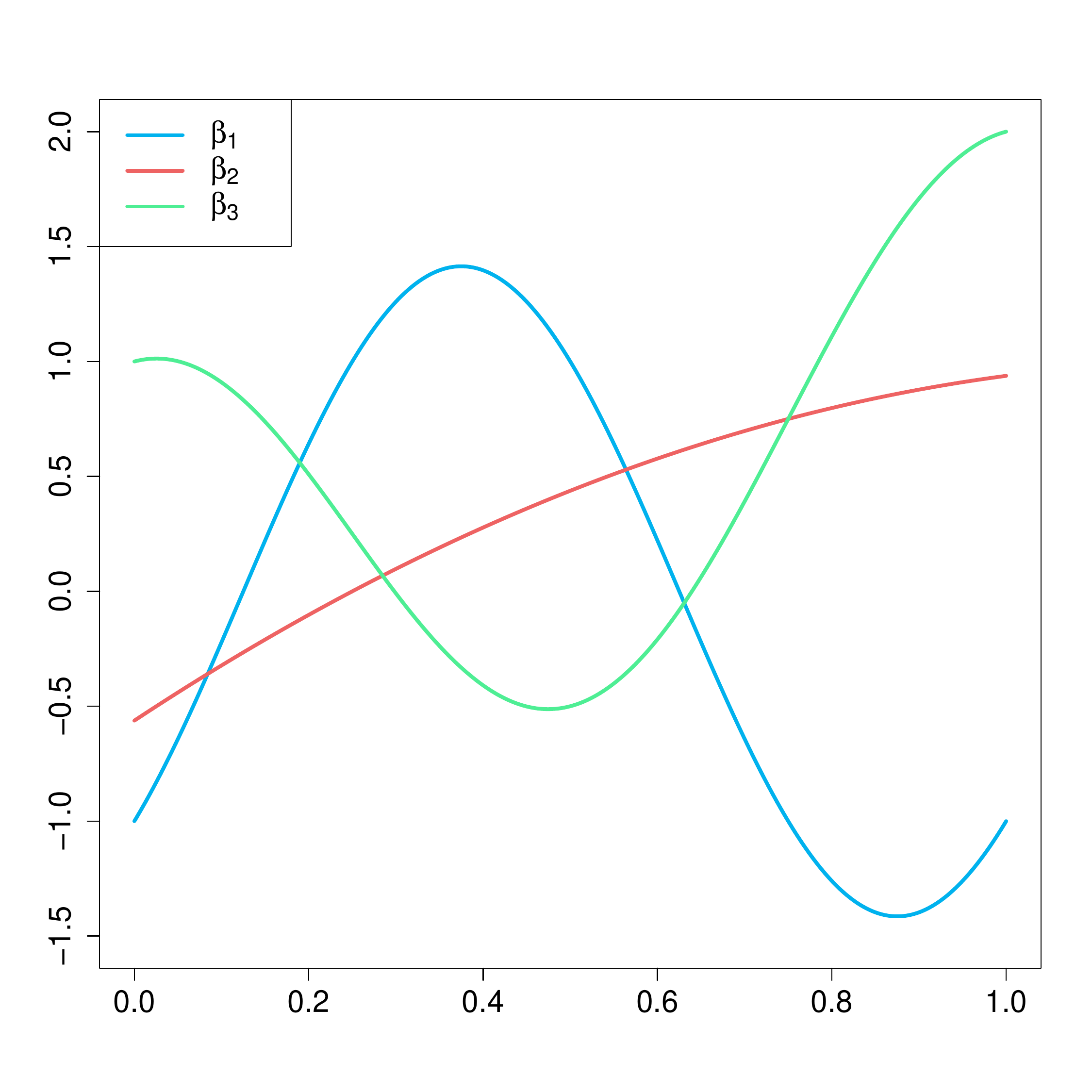}
	\caption{\small Functional slopes corresponding to the FLMSR considered in the Monte Carlo experiments.}
	\label{fig:Betas-Model-1}
\end{figure}

For each functional slope and a sample size $n=50,100,200$, $M=1000$ samples of $n$ independent random pairs from the FLMSR are generated and $\beta$ is estimated with the FPC estimator in \cite{Cardot2007} and the LASSO-selected estimator in \cite{Garcia-Portugues2020}. These estimators of $\beta$ obtained with the fully observed samples are used as benchmarks for the estimators in samples with missing responses. Next, in each generated sample, missing responses are generated with the observance probability
\begin{align}\label{Obs-Prob}
	p\left(\xcal\right)=\frac{1}{1+\exp\left(-\eta\|\xcal\|^2\right)},
\end{align}
where $\eta=0.5,1,2$ is such that the expected percentages of missing responses are approximately $35\%$, $27\%$, and $20\%$, respectively. Consequently, predictors with smaller norms have a greater probability of generating missing responses. In particular, the simulation studies in \cite{Ferraty2013}, \cite{Ling2015}, \cite{Ling2016}, \cite{Crambes2018}, and \cite{Febrero-Bande2019}, among others, considered $p\left(\cdot\right)$ in~\eqref{Obs-Prob} with $\eta=1$. Subsequently, in each sample with missing responses, $\beta$ is estimated with the six estimators proposed in Section~\ref{sec:Estimation-FLM-MAR}. For that, the upper bound $K_{\max}$ is increased until the last FPC is able to explain at most the $0.5\%$ of the variability of the predictors, a choice that gives rise to values of $K_{\max}$ between $4$ and $6$. The focus is on two aspects of these estimators: (\textit{i}) the Mean Square Error of Estimation (MSEE) of the eight estimates of $\beta$, two computed with the full samples and six computed with the samples with missing responses, given by $\mathbb{E}\big[\big\|\beta-\widehat{\beta}\big\|^2\big]$, that is estimated with the sample mean of the empirical MSEEs obtained from the $M=1000$ estimates; and (\textit{ii}) the time in seconds required to obtain each of the estimators. 

\begin{table}[hptb!]
	\centering
	\begin{tabular}{ccccc}
		& $\delta_0=0.00$ & $\delta_1=0.01$ & $\delta_2=0.02$ & $\delta_3=0.03$ \\ \midrule
		$\beta_{1}$ & \multicolumn{1}{|c}{$0.8232$} & $0.8244$ & $0.8285$ & $0.8334$ \\
		$\beta_{2}$ & \multicolumn{1}{|c}{$0.9490$} & $0.9491$ & $0.9495$ & $0.9500$ \\
		$\beta_{3}$ & \multicolumn{1}{|c}{$0.9709$} & $0.9710$ & $0.9712$ & $0.9713$  \\ \midrule
	\end{tabular}
	\caption{\small Coefficients of determination}
	\label{tab:Coefficients_Determination}
\end{table}

Given the wide range of results obtained, to save space only those cases that best illustrate the most representative conclusions of the results are reported. Figure~\ref{fig:log-MSEEs} shows the logarithm of the estimated MSEEs of the eight estimators for the functional slope $\beta_{3}$, observance probability parameters $\eta=0.5,2$, and sample sizes $n=50,200$. Three comments are in order. First, as expected, the reference estimators obtained with the full samples, denoted by $\widehat{\beta}_{\mathcal{C}}$ and $\widehat{\beta}_{\mathcal{CL}}$, have lower MSEEs than the estimators obtained with the samples with missing responses. However, the differences between the MSEEs of $\widehat{\beta}_{\mathcal{C}}$ and $\widehat{\beta}_{\mathcal{CL}}$ with respect to the imputed estimators $\widehat{\beta}_{\mathcal{I}}$ and $\widehat{\beta}_{\mathcal{IL}}$ and the inverse probability weighted estimators $\widehat{\beta}_{\mathcal{W}}$ and $\widehat{\beta}_{\mathcal{WL}}$ are very small when $\eta=2$, i.e., when the number of missing responses is small. Second, as also expected, the imputed and inverse probability weighted estimators $\widehat{\beta}_{\mathcal{I}}$, $\widehat{\beta}_{\mathcal{IL}}$, $\widehat{\beta}_{\mathcal{W}}$, and $\widehat{\beta}_{\mathcal{WL}}$ have lower estimated MSSEs than the simplified estimators $\widehat{\beta}_{\mathcal{S}}$ and $\widehat{\beta}_{\mathcal{SL}}$. Consequently, the imputation of responses with the simplified estimator is relevant for the estimation of $\beta$. Third, there is no consensus as to whether or not OLS+CV-based estimators are better than LASSO-based estimators in this setup, since the MSEEs of both cases are quite similar.

Figure~\ref{fig:log-Times} shows the logarithm of the time (in seconds) needed to compute these estimators of $\beta$ for the same setting as in Figure \ref{fig:log-MSEEs}. Three comments are in order. First, as might be expected, simplified estimators are the fastest to compute, since they only use pairs with observed responses. However, as seen above, imputed estimators perform better than simplified estimators at the cost of slightly increasing the computational cost. Second, estimators based on OLS+CV are computationally less demanding than LASSO-selected estimators, without reducing their estimated MSEEs. Third, the inverse probability weighted estimators $\widehat{\beta}_{\mathcal{W}}$ and $\widehat{\beta}_{\mathcal{WL}}$ are the most computationally expensive, which is expected given that they require to compute the Nadaraya--Watson estimator of the MAR mechanism $p\left(\cdot\right)$ in \eqref{Estimator-NW}. The boxplots of the MSSEs and the times corresponding to the remaining combinations of functional slope, observance probability parameter, and sample size do not provide different conclusions from those given for Figures \ref{fig:log-MSEEs} and \ref{fig:log-Times}.

\begin{figure}[!h]
	\centering
	\includegraphics[width=.49\textwidth,clip,trim={0cm 0.5cm 0cm 1.25cm}]{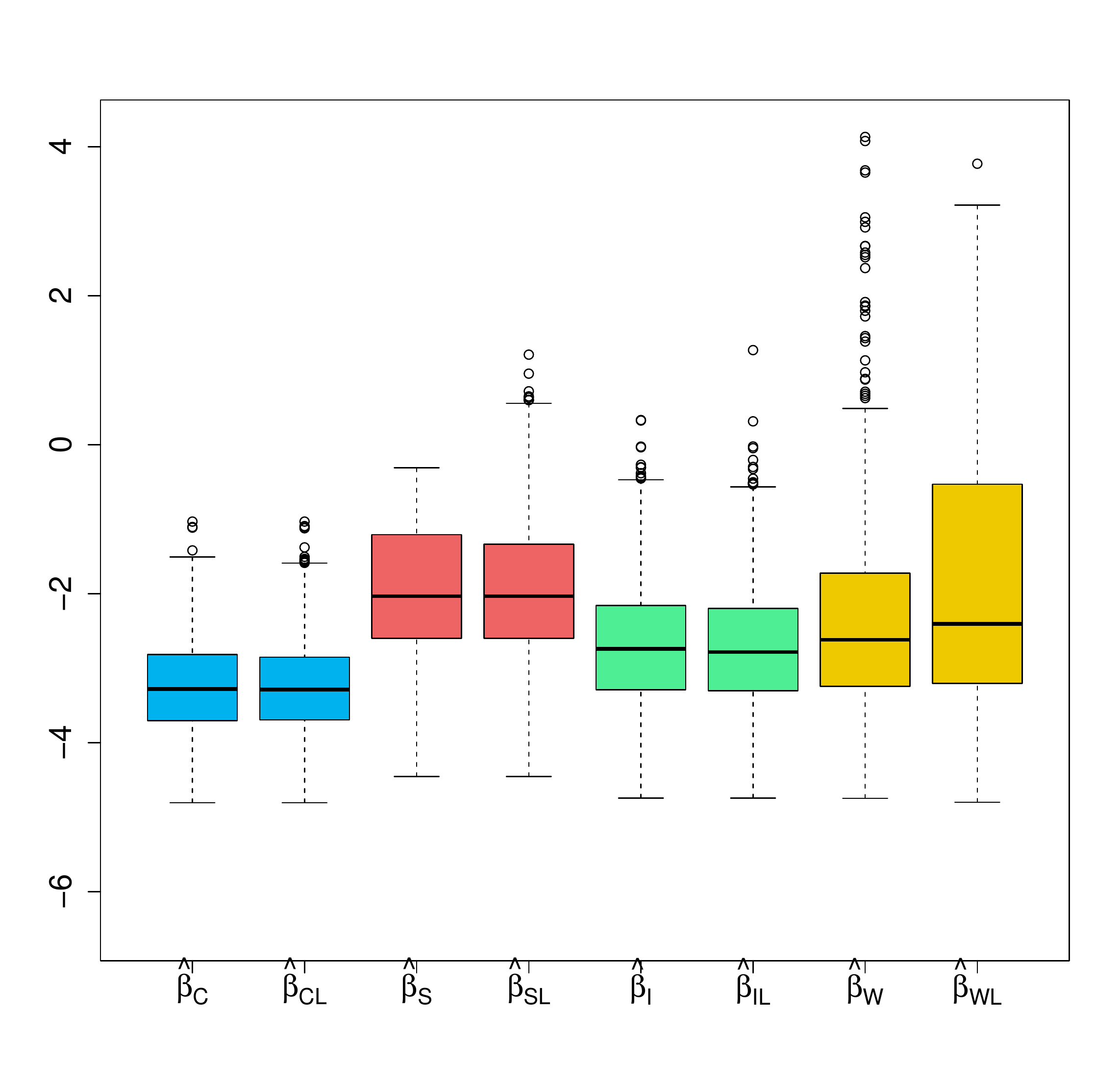}\includegraphics[width=.49\textwidth,clip,trim={0cm 0.5cm 0cm 1.25cm}]{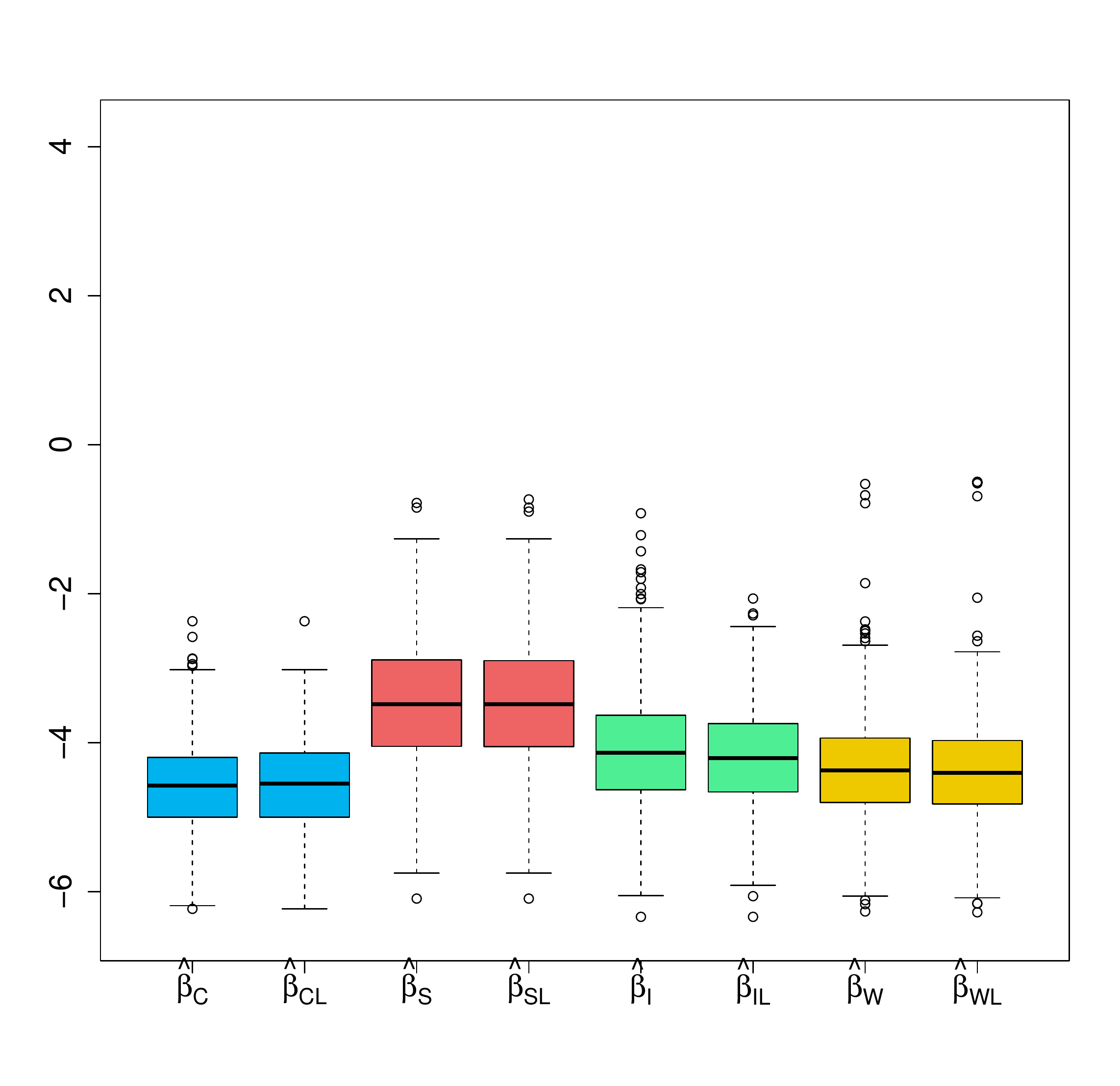}\\
	\includegraphics[width=.49\textwidth,clip,trim={0cm 0.5cm 0cm 1.25cm}]{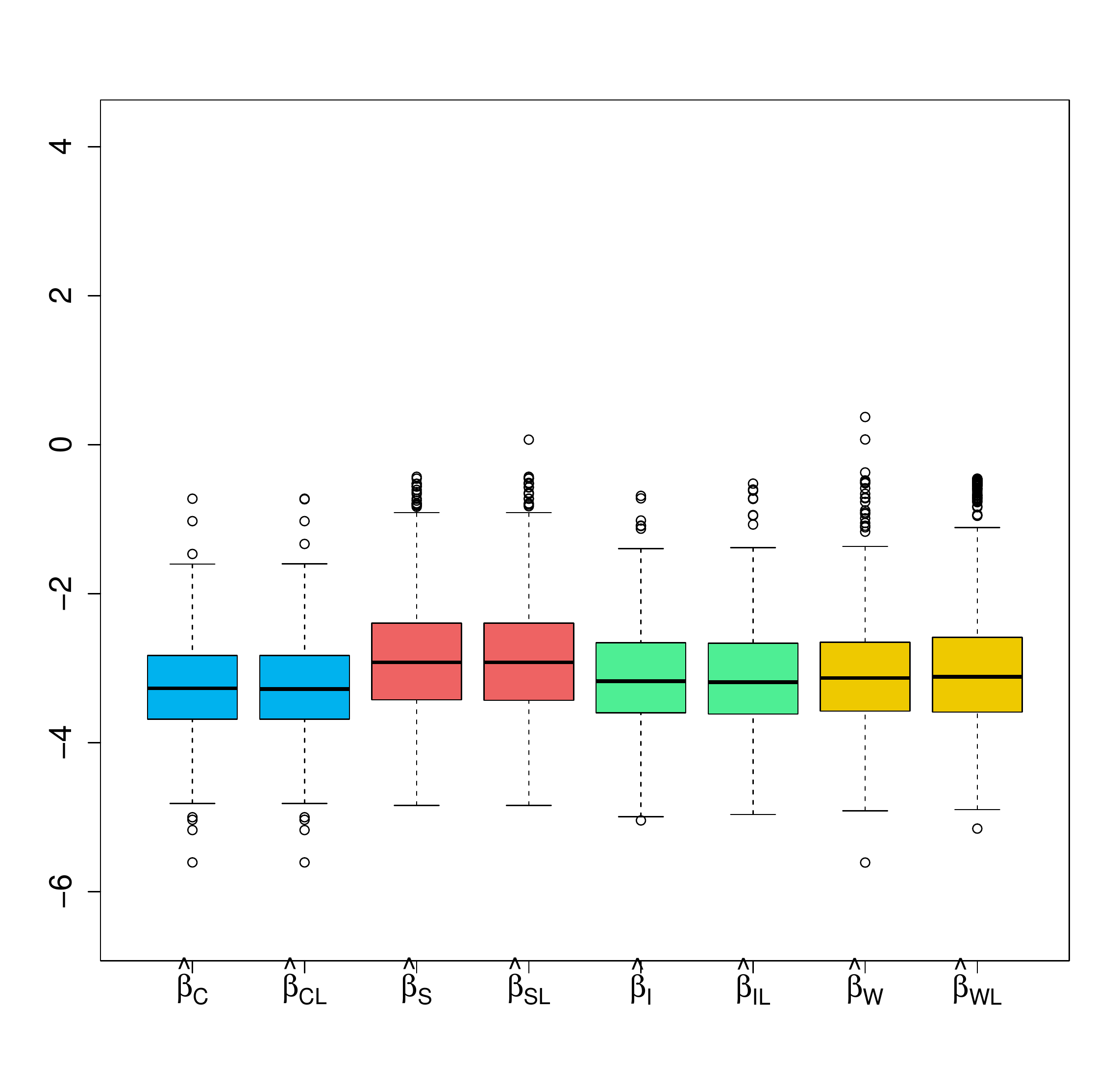}\includegraphics[width=.49\textwidth,clip,trim={0cm 0.5cm 0cm 1.25cm}]{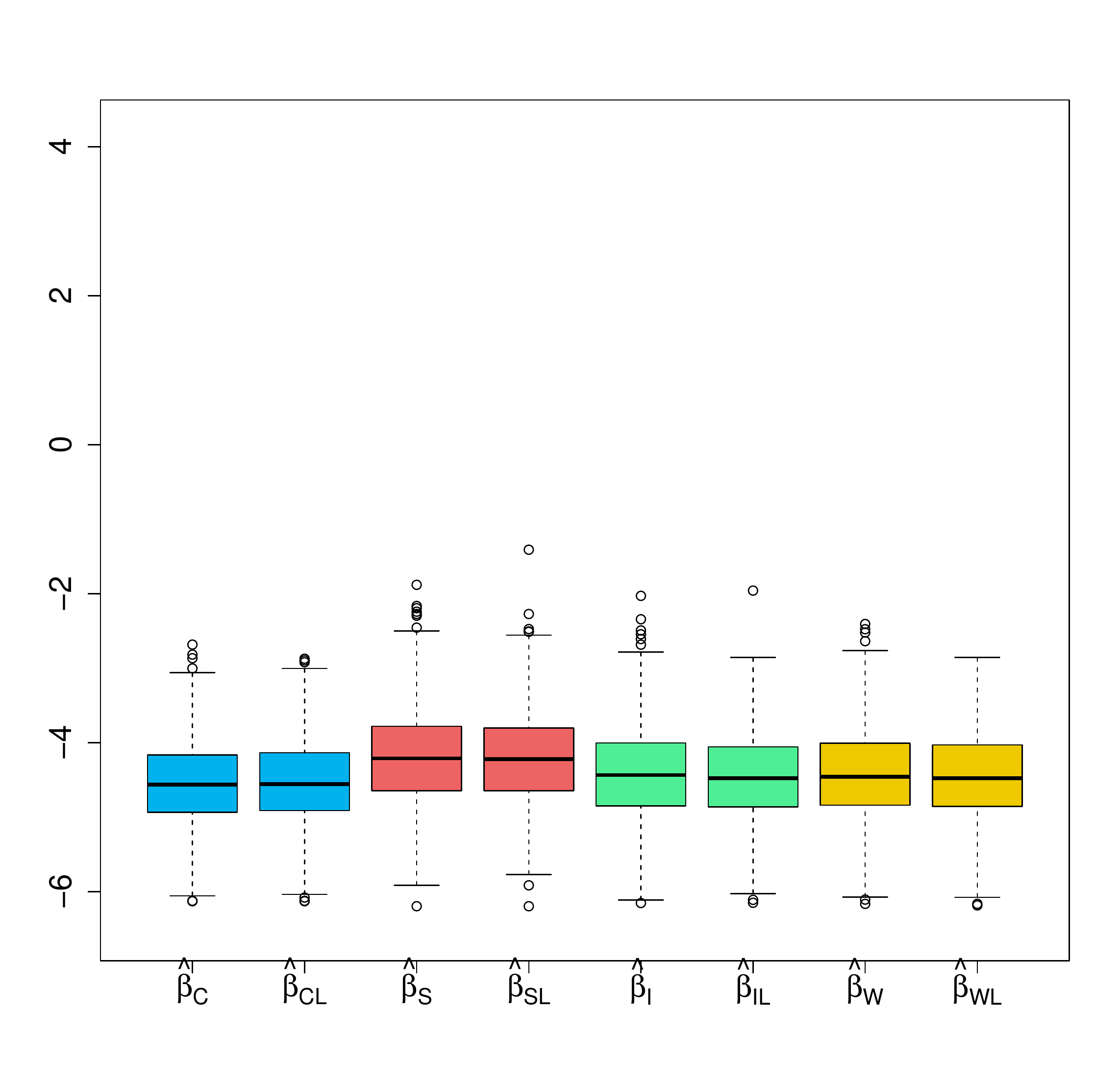}
	\caption{\small Logarithms of MSEEs for the eight estimators of $\beta_3$, for $n=50,200$ (columns, from left to right) and $\eta=0.5,2$ (rows, from top to bottom).}
	\label{fig:log-MSEEs}
\end{figure}

\begin{figure}[!h]
	\centering
	\includegraphics[width=.49\textwidth,clip,trim={0cm 0.5cm 0cm 1.25cm}]{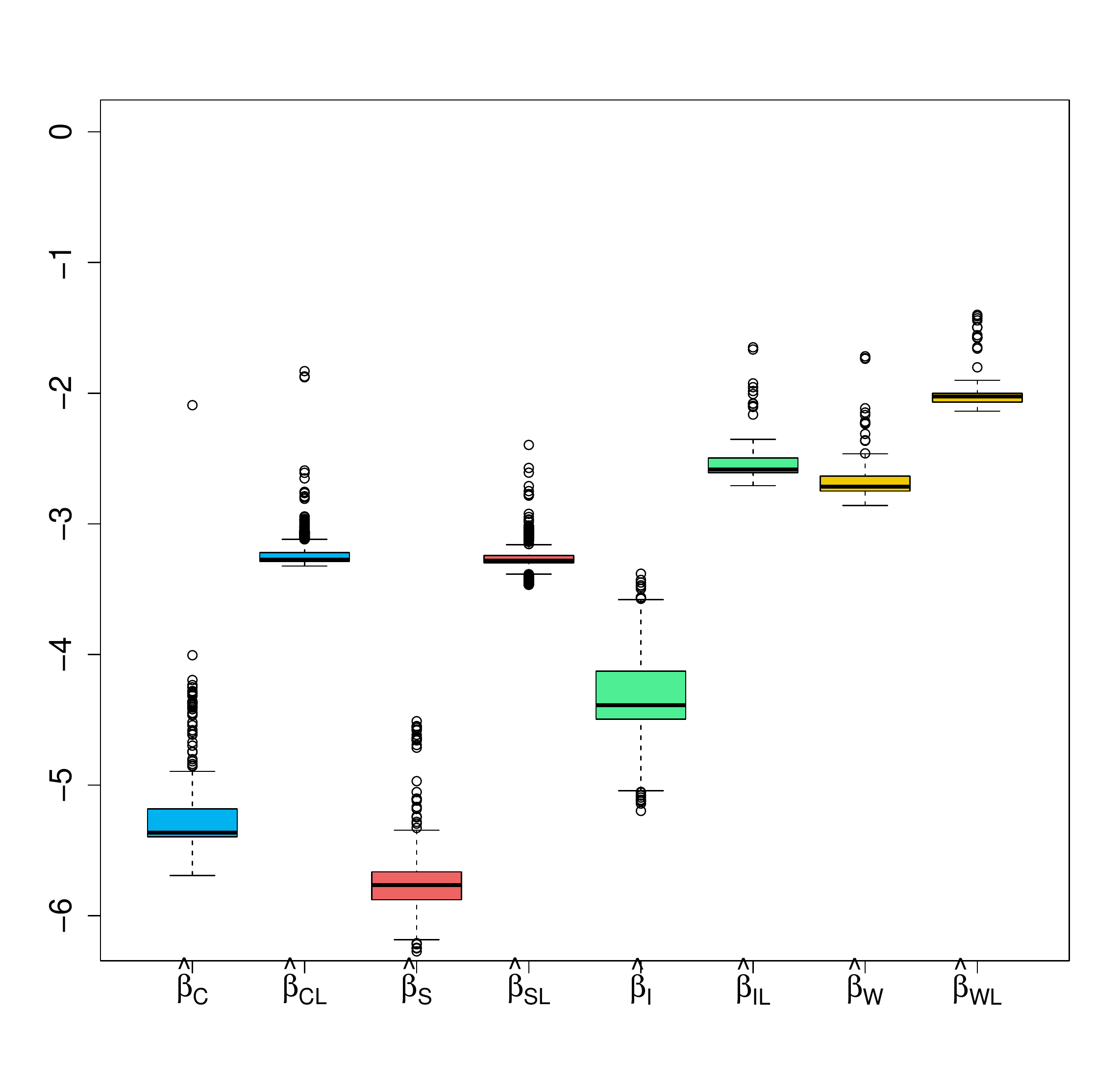}\includegraphics[width=.49\textwidth,clip,trim={0cm 0.5cm 0cm 1.25cm}]{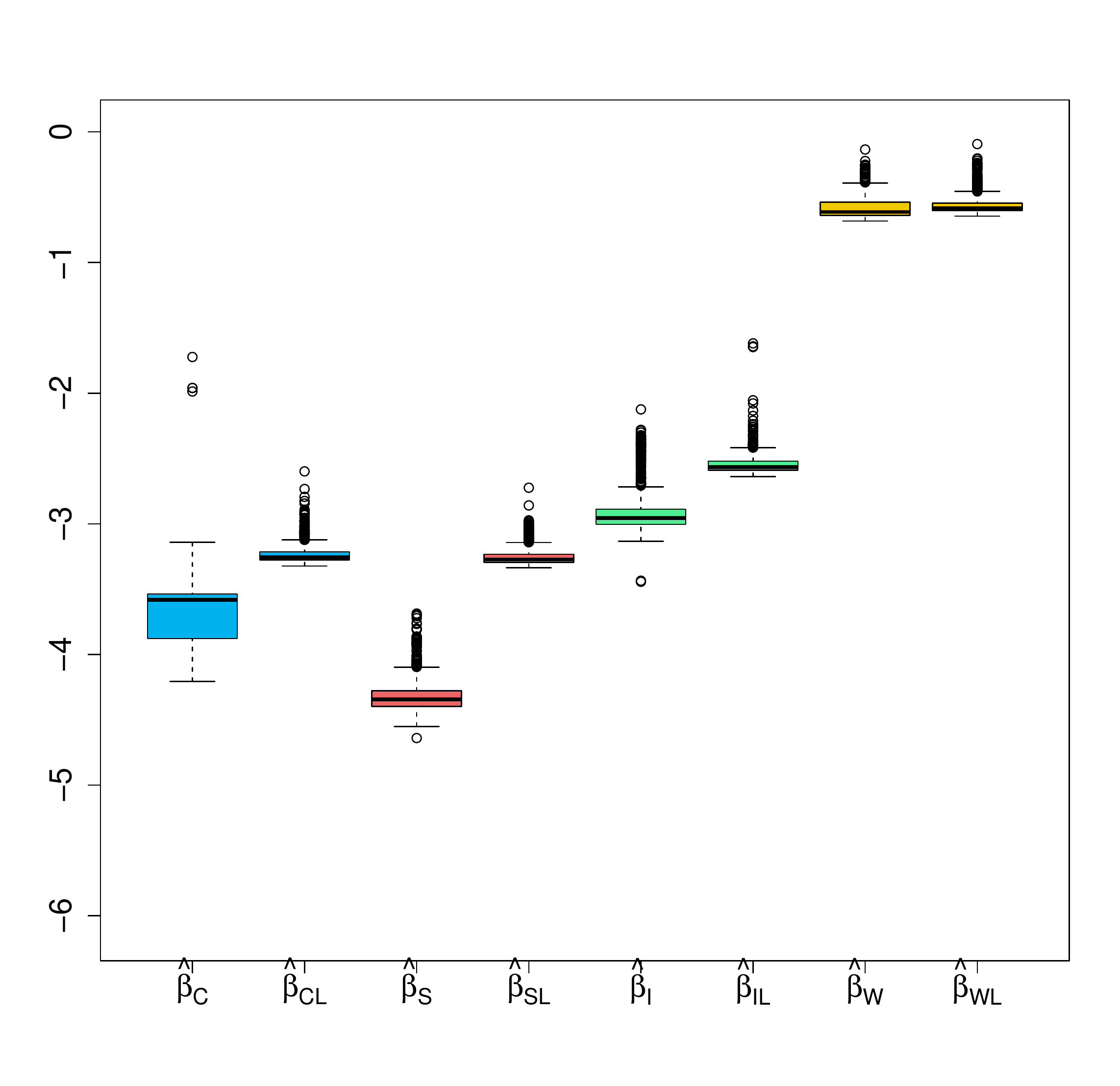}\\
	\includegraphics[width=.49\textwidth,clip,trim={0cm 0.5cm 0cm 1.25cm}]{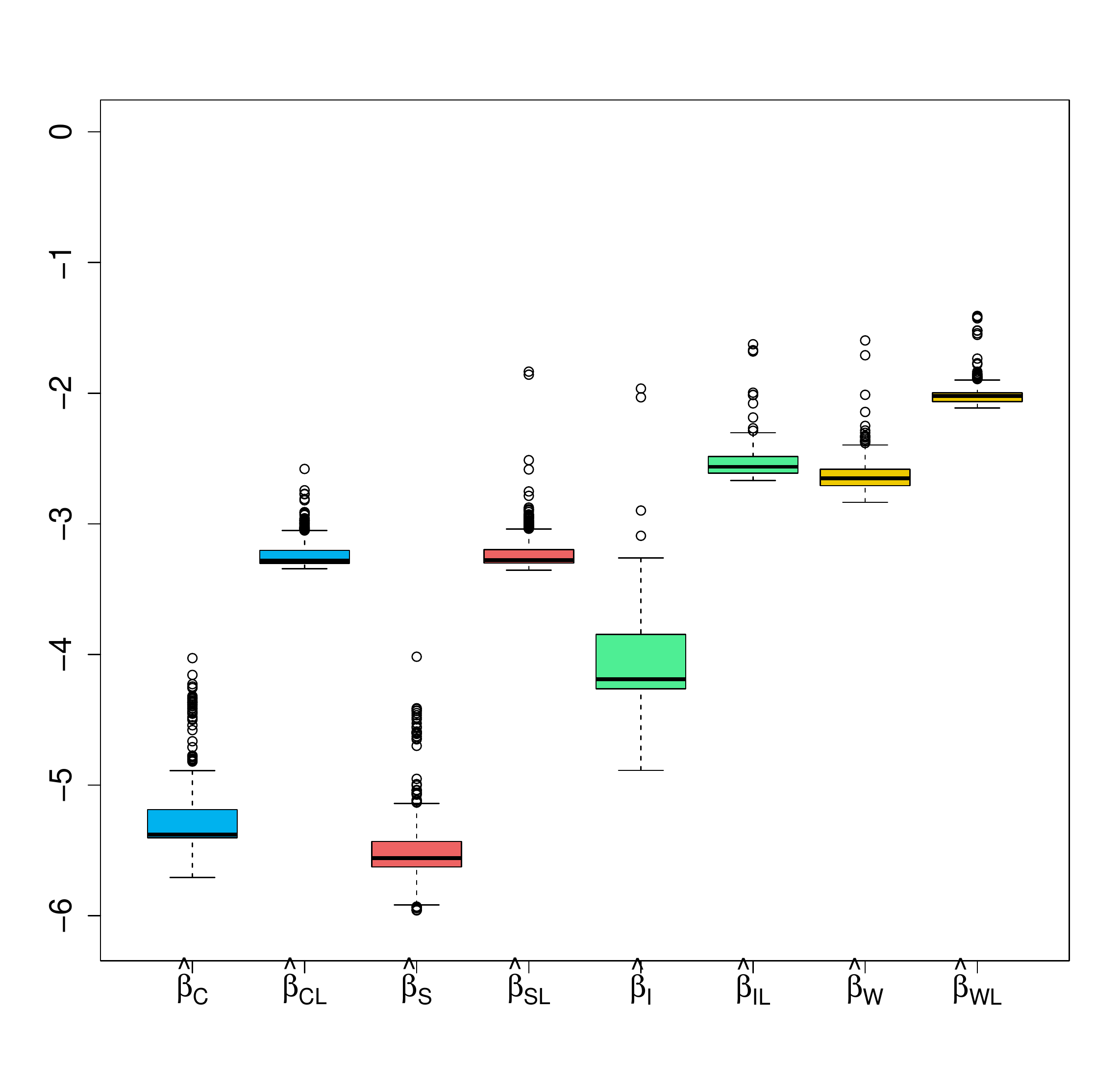}\includegraphics[width=.49\textwidth,clip,trim={0cm 0.5cm 0cm 1.25cm}]{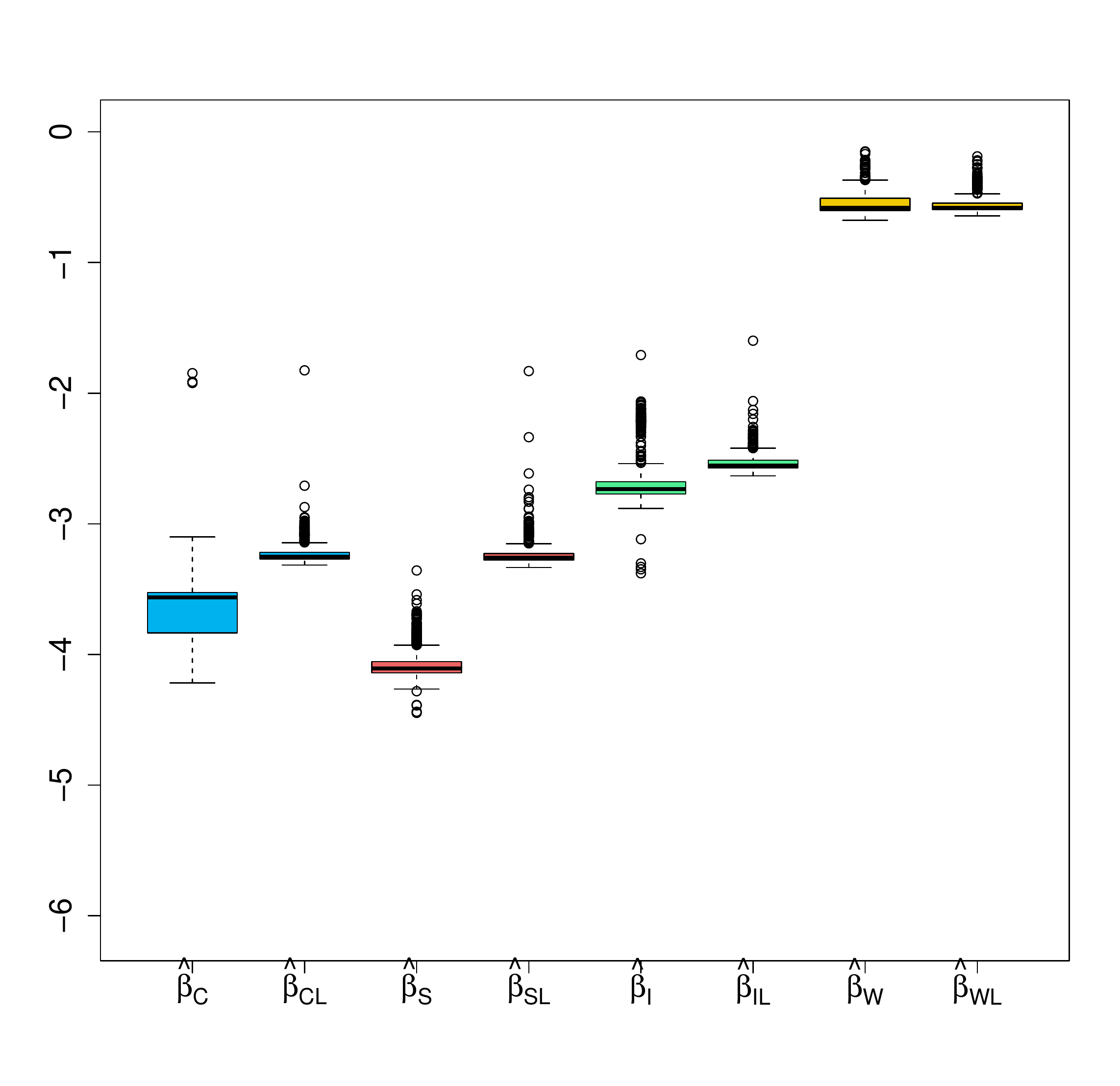}
	\caption{\small Logarithms of times (in seconds) needed to compute the eight estimators of $\beta_3$, for $n=50,200$ (columns, from left to right) and $\eta=0.5,2$ (rows, from top to bottom).}
	\label{fig:log-Times}
\end{figure}

\begin{table}[!t]
	\centering
	\begin{tabular}{cccccccccc}
		&  & \multicolumn{8}{c}{$\eta=0.5$} \\ \midrule
		$n$ & $\delta $ & $\widehat{\beta }_{\mathcal{C}}$ & $\widehat{\beta }_{\mathcal{CL}}$ & $\widehat{\beta }_{\mathcal{S}}$ & $\widehat{\beta }_{\mathcal{SL}}$ & $\widehat{\beta }_{\mathcal{I}}$ & $\widehat{\beta }_{\mathcal{IL}}$ & $\widehat{\beta }_{\mathcal{W}}$ & $\widehat{\beta }_{\mathcal{WL}}$ \\
		\midrule
		50 & \multicolumn{1}{|c}{0.00} & \multicolumn{1}{|c}{0.065} & 0.064 & 0.053
		& 0.054 & 0.059 & 0.058 & 0.046 & 0.035 \\
		& \multicolumn{1}{|c}{0.01} & \multicolumn{1}{|c}{0.161} & 0.163 & 0.142 &
		0.148 & 0.156 & 0.159 & 0.145 & 0.137 \\
		& \multicolumn{1}{|c}{0.02} & \multicolumn{1}{|c}{0.412} & 0.401 & 0.316 &
		0.306 & 0.369 & 0.348 & 0.343 & 0.292 \\
		& \multicolumn{1}{|c}{0.03} & \multicolumn{1}{|c}{0.666} & 0.651 & 0.540 &
		0.512 & 0.598 & 0.556 & 0.549 & 0.475 \\ \midrule
		100 & \multicolumn{1}{|c}{0.00} & \multicolumn{1}{|c}{0.055} & 0.055 & 0.056
		& 0.051 & 0.060 & 0.054 & 0.059 & 0.051 \\
		& \multicolumn{1}{|c}{0.01} & \multicolumn{1}{|c}{0.254} & 0.255 & 0.224 &
		0.219 & 0.247 & 0.239 & 0.249 & 0.228 \\
		& \multicolumn{1}{|c}{0.02} & \multicolumn{1}{|c}{0.727} & 0.715 & 0.640 &
		0.629 & 0.666 & 0.650 & 0.685 & 0.647 \\
		& \multicolumn{1}{|c}{0.03} & \multicolumn{1}{|c}{0.951} & 0.944 & 0.900 &
		0.900 & 0.916 & 0.913 & 0.909 & 0.879 \\ \midrule
		200 & \multicolumn{1}{|c}{0.00} & \multicolumn{1}{|c}{0.056} & 0.055 & 0.053
		& 0.055 & 0.056 & 0.056 & 0.054 & 0.052 \\
		& \multicolumn{1}{|c}{0.01} & \multicolumn{1}{|c}{0.462} & 0.461 & 0.458 &
		0.453 & 0.475 & 0.469 & 0.479 & 0.479 \\
		& \multicolumn{1}{|c}{0.02} & \multicolumn{1}{|c}{0.964} & 0.963 & 0.937 &
		0.931 & 0.947 & 0.937 & 0.951 & 0.932 \\
		& \multicolumn{1}{|c}{0.03} & \multicolumn{1}{|c}{0.999} & 0.999 & 0.999 &
		0.998 & 0.998 & 0.997 & 0.994 & 0.994 \\ \midrule
		&  & \multicolumn{8}{c}{$\eta=1$} \\ \midrule
		50 & \multicolumn{1}{|c}{0.00} & \multicolumn{1}{|c}{0.072} & 0.069 & 0.059
		& 0.055 & 0.067 & 0.061 & 0.064 & 0.060 \\
		& \multicolumn{1}{|c}{0.01} & \multicolumn{1}{|c}{0.150} & 0.148 & 0.143 &
		0.147 & 0.167 & 0.162 & 0.163 & 0.137 \\
		& \multicolumn{1}{|c}{0.02} & \multicolumn{1}{|c}{0.394} & 0.380 & 0.372 &
		0.359 & 0.402 & 0.378 & 0.398 & 0.351 \\
		& \multicolumn{1}{|c}{0.03} & \multicolumn{1}{|c}{0.671} & 0.645 & 0.618 &
		0.599 & 0.670 & 0.632 & 0.666 & 0.585 \\ \midrule
		100 & \multicolumn{1}{|c}{0.00} & \multicolumn{1}{|c}{0.057} & 0.055 & 0.055
		& 0.053 & 0.058 & 0.057 & 0.058 & 0.058 \\
		& \multicolumn{1}{|c}{0.01} & \multicolumn{1}{|c}{0.259} & 0.254 & 0.254 &
		0.251 & 0.266 & 0.260 & 0.279 & 0.265 \\
		& \multicolumn{1}{|c}{0.02} & \multicolumn{1}{|c}{0.719} & 0.707 & 0.731 &
		0.720 & 0.746 & 0.726 & 0.751 & 0.730 \\
		& \multicolumn{1}{|c}{0.03} & \multicolumn{1}{|c}{0.947} & 0.946 & 0.934 &
		0.932 & 0.943 & 0.942 & 0.944 & 0.939 \\ \midrule
		200 & \multicolumn{1}{|c}{0.00} & \multicolumn{1}{|c}{0.056} & 0.058 & 0.057
		& 0.060 & 0.063 & 0.066 & 0.065 & 0.066 \\
		& \multicolumn{1}{|c}{0.01} & \multicolumn{1}{|c}{0.479} & 0.477 & 0.509 &
		0.495 & 0.518 & 0.511 & 0.525 & 0.506 \\
		& \multicolumn{1}{|c}{0.02} & \multicolumn{1}{|c}{0.968} & 0.966 & 0.970 &
		0.967 & 0.977 & 0.974 & 0.978 & 0.971 \\
		& \multicolumn{1}{|c}{0.03} & \multicolumn{1}{|c}{1.000} & 1.000 & 1.000 &
		1.000 & 1.000 & 1.000 & 1.000 & 0.999 \\ \midrule
		&  & \multicolumn{8}{c}{$\eta=2$} \\ \midrule
		50 & \multicolumn{1}{|c}{0.00} & \multicolumn{1}{|c}{0.046} & 0.044 & 0.055
		& 0.059 & 0.055 & 0.060 & 0.056 & 0.057 \\
		& \multicolumn{1}{|c}{0.01} & \multicolumn{1}{|c}{0.142} & 0.136 & 0.156 &
		0.155 & 0.176 & 0.169 & 0.175 & 0.166 \\
		& \multicolumn{1}{|c}{0.02} & \multicolumn{1}{|c}{0.385} & 0.378 & 0.402 &
		0.392 & 0.419 & 0.405 & 0.419 & 0.385 \\
		& \multicolumn{1}{|c}{0.03} & \multicolumn{1}{|c}{0.674} & 0.655 & 0.676 &
		0.659 & 0.695 & 0.672 & 0.697 & 0.652 \\ \midrule
		100 & \multicolumn{1}{|c}{0.00} & \multicolumn{1}{|c}{0.063} & 0.057 & 0.057
		& 0.056 & 0.055 & 0.053 & 0.057 & 0.053 \\
		& \multicolumn{1}{|c}{0.01} & \multicolumn{1}{|c}{0.255} & 0.248 & 0.270 &
		0.270 & 0.286 & 0.279 & 0.291 & 0.278 \\
		& \multicolumn{1}{|c}{0.02} & \multicolumn{1}{|c}{0.741} & 0.712 & 0.758 &
		0.749 & 0.777 & 0.754 & 0.783 & 0.753 \\
		& \multicolumn{1}{|c}{0.03} & \multicolumn{1}{|c}{0.952} & 0.946 & 0.957 &
		0.952 & 0.959 & 0.954 & 0.959 & 0.957 \\ \midrule
		200 & \multicolumn{1}{|c}{0.00} & \multicolumn{1}{|c}{0.052} & 0.052 & 0.049
		& 0.049 & 0.057 & 0.056 & 0.054 & 0.053 \\
		& \multicolumn{1}{|c}{0.01} & \multicolumn{1}{|c}{0.512} & 0.508 & 0.558 &
		0.556 & 0.559 & 0.559 & 0.568 & 0.551 \\
		& \multicolumn{1}{|c}{0.02} & \multicolumn{1}{|c}{0.963} & 0.965 & 0.980 &
		0.977 & 0.983 & 0.979 & 0.984 & 0.976 \\
		& \multicolumn{1}{|c}{0.03} & \multicolumn{1}{|c}{1.000} & 1.000 & 1.000 &
		1.000 & 1.000 & 1.000 & 1.000 & 1.000 \\ \midrule
	\end{tabular}
	\caption{\small Rejection frequencies of the null hypothesis for significance level $\alpha=0.05$ for $\beta_1$.}
	\label{tab:Rejection_Frequencies_FLMSR_1}
\end{table}

\begin{table}[!b]
	\centering
	\begin{tabular}{cccccccccc}
		&  & \multicolumn{8}{c}{$\eta=0.5$} \\ \midrule
		$n$ & $\delta $ & $\widehat{\beta }_{\mathcal{C}}$ & $\widehat{\beta }_{\mathcal{CL}}$ & $\widehat{\beta }_{\mathcal{S}}$ & $\widehat{\beta }_{\mathcal{SL}}$ & $\widehat{\beta }_{\mathcal{I}}$ & $\widehat{\beta }_{\mathcal{IL}}$ & $\widehat{\beta }_{\mathcal{W}}$ & $\widehat{\beta }_{\mathcal{WL}}$ \\
		\midrule
		50 & \multicolumn{1}{|c}{0.00} & \multicolumn{1}{|c}{0.051} & 0.051 & 0.047
		& 0.049 & 0.052 & 0.051 & 0.050 & 0.037 \\
		& \multicolumn{1}{|c}{0.01} & \multicolumn{1}{|c}{0.171} & 0.170 & 0.124 &
		0.119 & 0.151 & 0.146 & 0.137 & 0.126 \\
		& \multicolumn{1}{|c}{0.02} & \multicolumn{1}{|c}{0.386} & 0.388 & 0.303 &
		0.300 & 0.349 & 0.341 & 0.314 & 0.278 \\
		& \multicolumn{1}{|c}{0.03} & \multicolumn{1}{|c}{0.692} & 0.688 & 0.543 &
		0.545 & 0.596 & 0.591 & 0.552 & 0.504 \\ \midrule
		100 & \multicolumn{1}{|c}{0.00} & \multicolumn{1}{|c}{0.040} & 0.040 & 0.043
		& 0.043 & 0.049 & 0.049 & 0.057 & 0.047 \\
		& \multicolumn{1}{|c}{0.01} & \multicolumn{1}{|c}{0.269} & 0.256 & 0.219 &
		0.215 & 0.230 & 0.237 & 0.242 & 0.231 \\
		& \multicolumn{1}{|c}{0.02} & \multicolumn{1}{|c}{0.689} & 0.688 & 0.618 &
		0.620 & 0.663 & 0.667 & 0.682 & 0.662 \\
		& \multicolumn{1}{|c}{0.03} & \multicolumn{1}{|c}{0.961} & 0.965 & 0.899 &
		0.900 & 0.927 & 0.925 & 0.925 & 0.904 \\ \midrule
		200 & \multicolumn{1}{|c}{0.00} & \multicolumn{1}{|c}{0.044} & 0.044 & 0.053
		& 0.053 & 0.051 & 0.054 & 0.056 & 0.053 \\
		& \multicolumn{1}{|c}{0.01} & \multicolumn{1}{|c}{0.446} & 0.438 & 0.444 &
		0.443 & 0.463 & 0.461 & 0.463 & 0.464 \\
		& \multicolumn{1}{|c}{0.02} & \multicolumn{1}{|c}{0.947} & 0.947 & 0.937 &
		0.937 & 0.949 & 0.947 & 0.948 & 0.940 \\
		& \multicolumn{1}{|c}{0.03} & \multicolumn{1}{|c}{1.000} & 1.000 & 0.999 &
		0.998 & 1.000 & 0.999 & 0.998 & 0.997 \\ \midrule
		&  & \multicolumn{8}{c}{$\eta=1$} \\ \midrule
		50 & \multicolumn{1}{|c}{0.00} & \multicolumn{1}{|c}{0.059} & 0.059 & 0.052
		& 0.054 & 0.059 & 0.059 & 0.052 & 0.059 \\
		& \multicolumn{1}{|c}{0.01} & \multicolumn{1}{|c}{0.155} & 0.152 & 0.142 &
		0.143 & 0.151 & 0.153 & 0.150 & 0.145 \\
		& \multicolumn{1}{|c}{0.02} & \multicolumn{1}{|c}{0.415} & 0.412 & 0.393 &
		0.397 & 0.435 & 0.439 & 0.421 & 0.393 \\
		& \multicolumn{1}{|c}{0.03} & \multicolumn{1}{|c}{0.690} & 0.684 & 0.649 &
		0.640 & 0.673 & 0.675 & 0.674 & 0.641 \\ \midrule
		100 & \multicolumn{1}{|c}{0.00} & \multicolumn{1}{|c}{0.061} & 0.060 & 0.053
		& 0.053 & 0.057 & 0.055 & 0.053 & 0.052 \\
		& \multicolumn{1}{|c}{0.01} & \multicolumn{1}{|c}{0.247} & 0.248 & 0.252 &
		0.250 & 0.262 & 0.266 & 0.276 & 0.268 \\
		& \multicolumn{1}{|c}{0.02} & \multicolumn{1}{|c}{0.699} & 0.706 & 0.715 &
		0.718 & 0.730 & 0.741 & 0.744 & 0.738 \\
		& \multicolumn{1}{|c}{0.03} & \multicolumn{1}{|c}{0.956} & 0.957 & 0.957 &
		0.956 & 0.959 & 0.964 & 0.961 & 0.959 \\ \midrule
		200 & \multicolumn{1}{|c}{0.00} & \multicolumn{1}{|c}{0.038} & 0.044 & 0.050
		& 0.049 & 0.049 & 0.051 & 0.051 & 0.052 \\
		& \multicolumn{1}{|c}{0.01} & \multicolumn{1}{|c}{0.489} & 0.486 & 0.483 &
		0.481 & 0.495 & 0.494 & 0.496 & 0.499 \\
		& \multicolumn{1}{|c}{0.02} & \multicolumn{1}{|c}{0.963} & 0.960 & 0.967 &
		0.966 & 0.972 & 0.968 & 0.973 & 0.970 \\
		& \multicolumn{1}{|c}{0.03} & \multicolumn{1}{|c}{1.000} & 1.000 & 1.000 &
		1.000 & 1.000 & 1.000 & 1.000 & 1.000 \\ \midrule
		&  & \multicolumn{8}{c}{$\eta=2$} \\ \midrule
		50 & \multicolumn{1}{|c}{0.00} & \multicolumn{1}{|c}{0.058} & 0.056 & 0.056
		& 0.057 & 0.059 & 0.054 & 0.052 & 0.051 \\
		& \multicolumn{1}{|c}{0.01} & \multicolumn{1}{|c}{0.145} & 0.140 & 0.141 &
		0.141 & 0.145 & 0.144 & 0.149 & 0.139 \\
		& \multicolumn{1}{|c}{0.02} & \multicolumn{1}{|c}{0.440} & 0.445 & 0.475 &
		0.480 & 0.484 & 0.487 & 0.483 & 0.479 \\
		& \multicolumn{1}{|c}{0.03} & \multicolumn{1}{|c}{0.698} & 0.694 & 0.716 &
		0.714 & 0.730 & 0.719 & 0.739 & 0.718 \\ \midrule
		100 & \multicolumn{1}{|c}{0.00} & \multicolumn{1}{|c}{0.054} & 0.056 & 0.041
		& 0.040 & 0.042 & 0.040 & 0.041 & 0.039 \\
		& \multicolumn{1}{|c}{0.01} & \multicolumn{1}{|c}{0.276} & 0.272 & 0.297 &
		0.298 & 0.306 & 0.300 & 0.308 & 0.304 \\
		& \multicolumn{1}{|c}{0.02} & \multicolumn{1}{|c}{0.746} & 0.748 & 0.768 &
		0.768 & 0.774 & 0.774 & 0.780 & 0.775 \\
		& \multicolumn{1}{|c}{0.03} & \multicolumn{1}{|c}{0.948} & 0.949 & 0.956 &
		0.958 & 0.958 & 0.959 & 0.960 & 0.959 \\ \midrule
		200 & \multicolumn{1}{|c}{0.00} & \multicolumn{1}{|c}{0.047} & 0.045 & 0.054
		& 0.059 & 0.053 & 0.060 & 0.052 & 0.060 \\
		& \multicolumn{1}{|c}{0.01} & \multicolumn{1}{|c}{0.449} & 0.439 & 0.512 &
		0.519 & 0.524 & 0.519 & 0.520 & 0.520 \\
		& \multicolumn{1}{|c}{0.02} & \multicolumn{1}{|c}{0.952} & 0.949 & 0.966 &
		0.969 & 0.970 & 0.971 & 0.973 & 0.974 \\
		& \multicolumn{1}{|c}{0.03} & \multicolumn{1}{|c}{1.000} & 0.999 & 1.000 &
		1.000 & 1.000 & 1.000 & 1.000 & 1.000 \\ \midrule
	\end{tabular}
	\caption{\small Rejection frequencies of the null hypothesis for significance level $\alpha=0.05$ for $\beta_2$.}
	\label{tab:Rejection_Frequencies_FLMSR_2}
\end{table}

\begin{table}[!h]
	\centering
	\begin{tabular}{cccccccccc}
		&  & \multicolumn{8}{c}{$\eta=0.5$} \\ \midrule
		$n$ & $\delta $ & $\widehat{\beta }_{\mathcal{C}}$ & $\widehat{\beta }_{\mathcal{CL}}$ & $\widehat{\beta }_{\mathcal{S}}$ & $\widehat{\beta }_{\mathcal{SL}}$ & $\widehat{\beta }_{\mathcal{I}}$ & $\widehat{\beta }_{\mathcal{IL}}$ & $\widehat{\beta }_{\mathcal{W}}$ & $\widehat{\beta }_{\mathcal{WL}}$ \\
		\midrule
		50 & \multicolumn{1}{|c}{0.00} & \multicolumn{1}{|c}{0.064} & 0.068 & 0.039
		& 0.032 & 0.047 & 0.047 & 0.054 & 0.051 \\
		& \multicolumn{1}{|c}{0.01} & \multicolumn{1}{|c}{0.152} & 0.153 & 0.095 &
		0.088 & 0.136 & 0.126 & 0.126 & 0.104 \\
		& \multicolumn{1}{|c}{0.02} & \multicolumn{1}{|c}{0.416} & 0.419 & 0.252 &
		0.246 & 0.361 & 0.348 & 0.330 & 0.298 \\
		& \multicolumn{1}{|c}{0.03} & \multicolumn{1}{|c}{0.675} & 0.668 & 0.461 &
		0.434 & 0.583 & 0.543 & 0.532 & 0.467 \\ \midrule
		100 & \multicolumn{1}{|c}{0.00} & \multicolumn{1}{|c}{0.048} & 0.048 & 0.042
		& 0.041 & 0.052 & 0.055 & 0.059 & 0.053 \\
		& \multicolumn{1}{|c}{0.01} & \multicolumn{1}{|c}{0.265} & 0.262 & 0.176 &
		0.171 & 0.243 & 0.239 & 0.256 & 0.240 \\
		& \multicolumn{1}{|c}{0.02} & \multicolumn{1}{|c}{0.739} & 0.738 & 0.577 &
		0.577 & 0.667 & 0.674 & 0.678 & 0.650 \\
		& \multicolumn{1}{|c}{0.03} & \multicolumn{1}{|c}{0.949} & 0.949 & 0.853 &
		0.857 & 0.918 & 0.921 & 0.924 & 0.900 \\ \midrule
		200 & \multicolumn{1}{|c}{0.00} & \multicolumn{1}{|c}{0.045} & 0.046 & 0.036
		& 0.035 & 0.051 & 0.045 & 0.052 & 0.047 \\
		& \multicolumn{1}{|c}{0.01} & \multicolumn{1}{|c}{0.485} & 0.488 & 0.391 &
		0.390 & 0.448 & 0.447 & 0.464 & 0.463 \\
		& \multicolumn{1}{|c}{0.02} & \multicolumn{1}{|c}{0.950} & 0.946 & 0.904 &
		0.904 & 0.937 & 0.940 & 0.947 & 0.944 \\
		& \multicolumn{1}{|c}{0.03} & \multicolumn{1}{|c}{1.000} & 1.000 & 0.994 &
		0.995 & 0.999 & 0.999 & 0.993 & 0.989 \\ \midrule
		&  & \multicolumn{8}{c}{$\eta=1$} \\ \midrule
		50 & \multicolumn{1}{|c}{0.00} & \multicolumn{1}{|c}{0.061} & 0.063 & 0.045
		& 0.044 & 0.060 & 0.065 & 0.067 & 0.062 \\
		& \multicolumn{1}{|c}{0.01} & \multicolumn{1}{|c}{0.147} & 0.148 & 0.122 &
		0.120 & 0.149 & 0.145 & 0.151 & 0.145 \\
		& \multicolumn{1}{|c}{0.02} & \multicolumn{1}{|c}{0.411} & 0.410 & 0.335 &
		0.342 & 0.416 & 0.414 & 0.419 & 0.397 \\
		& \multicolumn{1}{|c}{0.03} & \multicolumn{1}{|c}{0.686} & 0.676 & 0.572 &
		0.569 & 0.694 & 0.663 & 0.678 & 0.614 \\ \midrule
		100 & \multicolumn{1}{|c}{0.00} & \multicolumn{1}{|c}{0.057} & 0.056 & 0.046
		& 0.045 & 0.059 & 0.057 & 0.055 & 0.058 \\
		& \multicolumn{1}{|c}{0.01} & \multicolumn{1}{|c}{0.260} & 0.259 & 0.261 &
		0.257 & 0.278 & 0.283 & 0.293 & 0.291 \\
		& \multicolumn{1}{|c}{0.02} & \multicolumn{1}{|c}{0.738} & 0.744 & 0.672 &
		0.668 & 0.722 & 0.721 & 0.739 & 0.727 \\
		& \multicolumn{1}{|c}{0.03} & \multicolumn{1}{|c}{0.958} & 0.954 & 0.922 &
		0.924 & 0.951 & 0.950 & 0.952 & 0.948 \\ \midrule
		200 & \multicolumn{1}{|c}{0.00} & \multicolumn{1}{|c}{0.046} & 0.059 & 0.042
		& 0.042 & 0.039 & 0.040 & 0.043 & 0.041 \\
		& \multicolumn{1}{|c}{0.01} & \multicolumn{1}{|c}{0.462} & 0.455 & 0.459 &
		0.458 & 0.501 & 0.498 & 0.495 & 0.497 \\
		& \multicolumn{1}{|c}{0.02} & \multicolumn{1}{|c}{0.961} & 0.960 & 0.956 &
		0.955 & 0.968 & 0.966 & 0.969 & 0.963 \\
		& \multicolumn{1}{|c}{0.03} & \multicolumn{1}{|c}{1.000} & 1.000 & 1.000 &
		1.000 & 1.000 & 1.000 & 1.000 & 1.000 \\ \midrule
		&  & \multicolumn{8}{c}{$\eta=2$} \\ \midrule
		50 & \multicolumn{1}{|c}{0.00} & \multicolumn{1}{|c}{0.058} & 0.060 & 0.056
		& 0.053 & 0.064 & 0.064 & 0.060 & 0.061 \\
		& \multicolumn{1}{|c}{0.01} & \multicolumn{1}{|c}{0.158} & 0.159 & 0.147 &
		0.144 & 0.177 & 0.173 & 0.181 & 0.176 \\
		& \multicolumn{1}{|c}{0.02} & \multicolumn{1}{|c}{0.406} & 0.400 & 0.378 &
		0.380 & 0.432 & 0.426 & 0.440 & 0.421 \\
		& \multicolumn{1}{|c}{0.03} & \multicolumn{1}{|c}{0.692} & 0.682 & 0.666 &
		0.662 & 0.715 & 0.699 & 0.721 & 0.690 \\ \midrule
		100 & \multicolumn{1}{|c}{0.00} & \multicolumn{1}{|c}{0.051} & 0.049 & 0.042
		& 0.041 & 0.049 & 0.050 & 0.047 & 0.048 \\
		& \multicolumn{1}{|c}{0.01} & \multicolumn{1}{|c}{0.275} & 0.271 & 0.261 &
		0.259 & 0.277 & 0.278 & 0.274 & 0.270 \\
		& \multicolumn{1}{|c}{0.02} & \multicolumn{1}{|c}{0.727} & 0.728 & 0.738 &
		0.735 & 0.762 & 0.760 & 0.763 & 0.760 \\
		& \multicolumn{1}{|c}{0.03} & \multicolumn{1}{|c}{0.942} & 0.945 & 0.945 &
		0.949 & 0.954 & 0.955 & 0.959 & 0.957 \\ \midrule
		200 & \multicolumn{1}{|c}{0.00} & \multicolumn{1}{|c}{0.049} & 0.050 & 0.063
		& 0.062 & 0.062 & 0.062 & 0.064 & 0.062 \\
		& \multicolumn{1}{|c}{0.01} & \multicolumn{1}{|c}{0.458} & 0.455 & 0.505 &
		0.507 & 0.518 & 0.516 & 0.521 & 0.519 \\
		& \multicolumn{1}{|c}{0.02} & \multicolumn{1}{|c}{0.962} & 0.961 & 0.967 &
		0.968 & 0.970 & 0.971 & 0.972 & 0.972 \\
		& \multicolumn{1}{|c}{0.03} & \multicolumn{1}{|c}{1.000} & 1.000 & 1.000 &
		1.000 & 1.000 & 1.000 & 1.000 & 1.000 \\ \midrule
	\end{tabular}
	\caption{\small Rejection frequencies of the null hypothesis for significance level $\alpha=0.05$ for $\beta_3$.}
	\label{tab:Rejection_Frequencies_FLMSR_3}
\end{table}

\subsection{Performance of the testing procedure}
\label{subsec:Performance}

The second Monte Carlo experiment analyzes the performance of the proposed testing procedure under the composite hypothesis \eqref{Null-Hypothesis}. For that, the following data-generating process is considered:
\begin{align}\label{FLMSR-delta}
	Y=\left\langle\Xcal,\beta_j\right\rangle+\delta_{k}\left\|\Xcal\right\|^2+\varepsilon,
\end{align}
where the functional predictor $\Xcal$, the functional slopes $\beta_j$, for $j=1,2,3$, and the error variance $\sigma_{\varepsilon}^2$ take the same value as in the first simulation experiment, while $\delta_k=0.01\cdot k$, for $k=0,1,2,3$, is a coefficient that measures the degree of deviation from the null hypothesis. Obviously, $\delta_{0}=0$ corresponds to the null hypothesis \eqref{Null-Hypothesis}, while increasingly larger values $\delta_{k}$ correspond to alternative hypotheses that are increasingly farther away from $H_{0}$. The resulting coefficients of determination are those in Table~\ref{tab:Coefficients_Determination} for each model and value of $\delta_{k}$. Note how little effect the nonlinear part of the FLMSRs has even for high values of $\delta_{k}$. This implies that detecting such nonlinearity can be challenging. The probability of observance, the parameters $\eta$, and the sample sizes $n$ are exactly the same as in the previous Monte Carlo experiment. For each combination of $\beta$, $\delta$, and $n$, $M=1000$ samples of $n$ independent random pairs from \eqref{FLMSR-delta} are generated and, for each sample, the testing procedure for the FPC estimator of $\beta$ in \cite{Cardot2007} and the LASSO-selected estimator in \cite{Garcia-Portugues2020} is run. Then, in each generated sample, missing responses are introduced with the observance probability \eqref{Obs-Prob} for one of the three values of $\eta$ considered and the proposed testing procedure for the six estimators of $\beta$ given in Section~\ref{sec:Estimation-FLM-MAR} is run. The $p$-values are computed with $B=1000$ bootstrap samples.

Tables~\ref{tab:Rejection_Frequencies_FLMSR_1}--\ref{tab:Rejection_Frequencies_FLMSR_3} show the rejection frequencies of the null hypothesis for the significance level $\alpha=0.05$ for the eight estimators of $\beta$, for the FLMSRs with $\beta_1$, $\beta_2$, and $\beta_3$, respectively. In each table, the rejection frequencies for the three values of $\eta$, the four values of $\delta_{k}$, and the three sample sizes $n$ are shown. Several comments are in order. First, the testing procedure seems to be able to calibrate the nominal size well, since there do not seem to be large deviations with respect to this value. The largest deviation is found for $\beta_1$, with $\eta=1$ and $n=50$, where the rejection frequency of the null hypothesis is $0.073$. Second, considering the small effect of the nonlinear part of the model even for $\delta_{3}$, the procedure, with and without missing responses, appears to be able to detect this nonlinearity in a large proportion of the generated samples, indicating that the performance of the procedure is satisfactory. Third, in the vast majority of cases, the frequency of rejection of the null hypothesis under the alternative is higher for estimators obtained with the full sample. It is true that for $\beta_1$ there are cases where paradoxically this is not the case, but this is most likely due to small-size distortions. Fourth, although the differences may sometimes seem small, there is a pattern that indicates that the imputed and inverse probability weighted estimators are more powerful than the simplified estimators. Of course, the larger the sample size, the smaller the difference between the estimators. However, larger differences can be found when the sample size is not large. For example, for $\beta_3$ (Table \ref{tab:Rejection_Frequencies_FLMSR_3}) with $n=50$, $\eta=0.5$, and $\delta=0.2$, the relative increase in the rejection frequency between the imputed and the simplified estimator is $43.25\%$, while with $n=100$, $\eta=0.5$, and $\delta=0.1$, the relative increase in rejection frequency between the inverse probability weighted and the simplified estimator is $45.45\%$. Fifth, as expected, the smaller the values of $\eta$ and/or $n$, the lower the frequency of rejection of estimators based on samples with missing responses. Sixth, there appear to be no major differences between the OLS+CV-based estimators and the LASSO-based estimators, so either can be used to test linearity. Arguably, the main conclusion is that there is a moderate loss in the power of the test if missing responses are simply discarded, especially when the sample size is not large. This is important, since discarding pairs of observations that include missing data in regression problems is a very common technique. Additionally, inverse probability weighted estimators could be expected to perform better than imputed estimators, taking into account the probability of generating missing responses. However, the behavior of all of them is quite similar, possibly because the observance probability in \eqref{Obs-Prob} must be estimated as a function of the sample.

\section{Real data example}
\label{sec:Real}

\cite{Febrero-Bande2019} analyzed a dataset in which the functional predictor is the mean curve of the annual average daily temperature observed in $73$ Spanish weather stations for the period 1980--2009 and the real response is the average of the number of sunny days per year, where a sunny day means that there are no opaque clouds for more than $80\%$ of daylight hours. Not all responses are recorded, as some weather stations do not measure this variable, possibly due to insufficient equipment to collect such data. A FLMSR was assumed, although no formal validation was given for this assumption. Here, the main objective of the analysis is to evaluate the linearity hypothesis and thus determine whether the conclusions presented in that article are well-founded. As in \cite{Febrero-Bande2019}, a group of temperature curves corresponding to stations in the Canary Islands and the Port of Navacerrada, one of the coldest locations in Spain, are excluded from the analysis because they are outlying with respect to most of the curves. The temperatures of the remaining $65$ stations after undergoing B-spline smoothing are shown in the left panel of Figure~\ref{fig:Sunny-Days-Data}. The temperatures corresponding to missing responses are shown in red and, as it can be observed, these curves are central in the sample. Therefore, the MAR assumption seems reasonable since the missing responses correspond to weather stations with mild temperatures. The right panel of Figure~\ref{fig:Sunny-Days-Data} shows the boxplot for the $50$ observed responses. The percentage of missing responses is $20.63\%$.

\begin{figure}[!ht]
	\centering
	\includegraphics[width=.495\textwidth,clip,trim={0cm 0.5cm 0cm 1.25cm}]{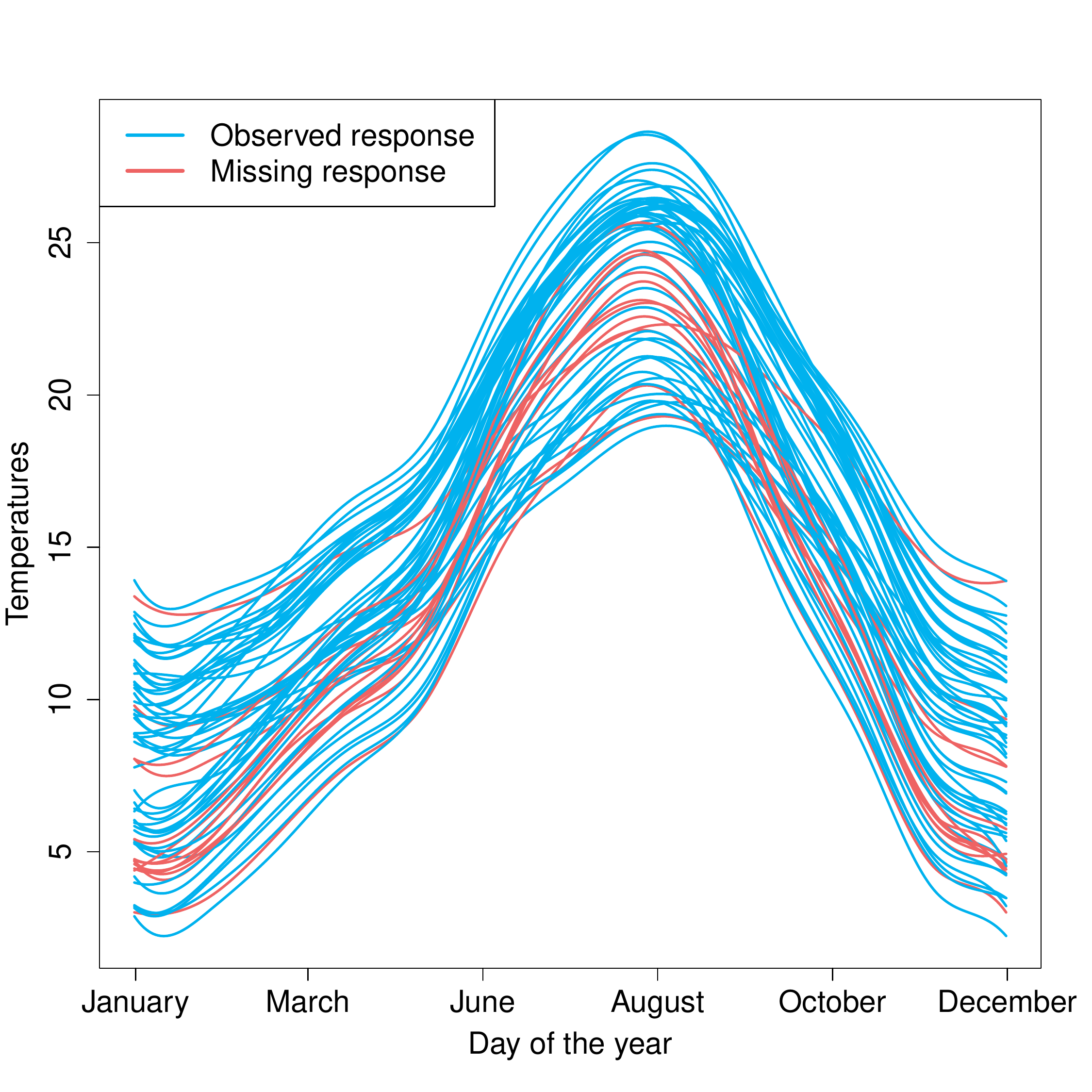}\includegraphics[width=.495\textwidth,clip,trim={0cm 0.5cm 0cm 1.25cm}]{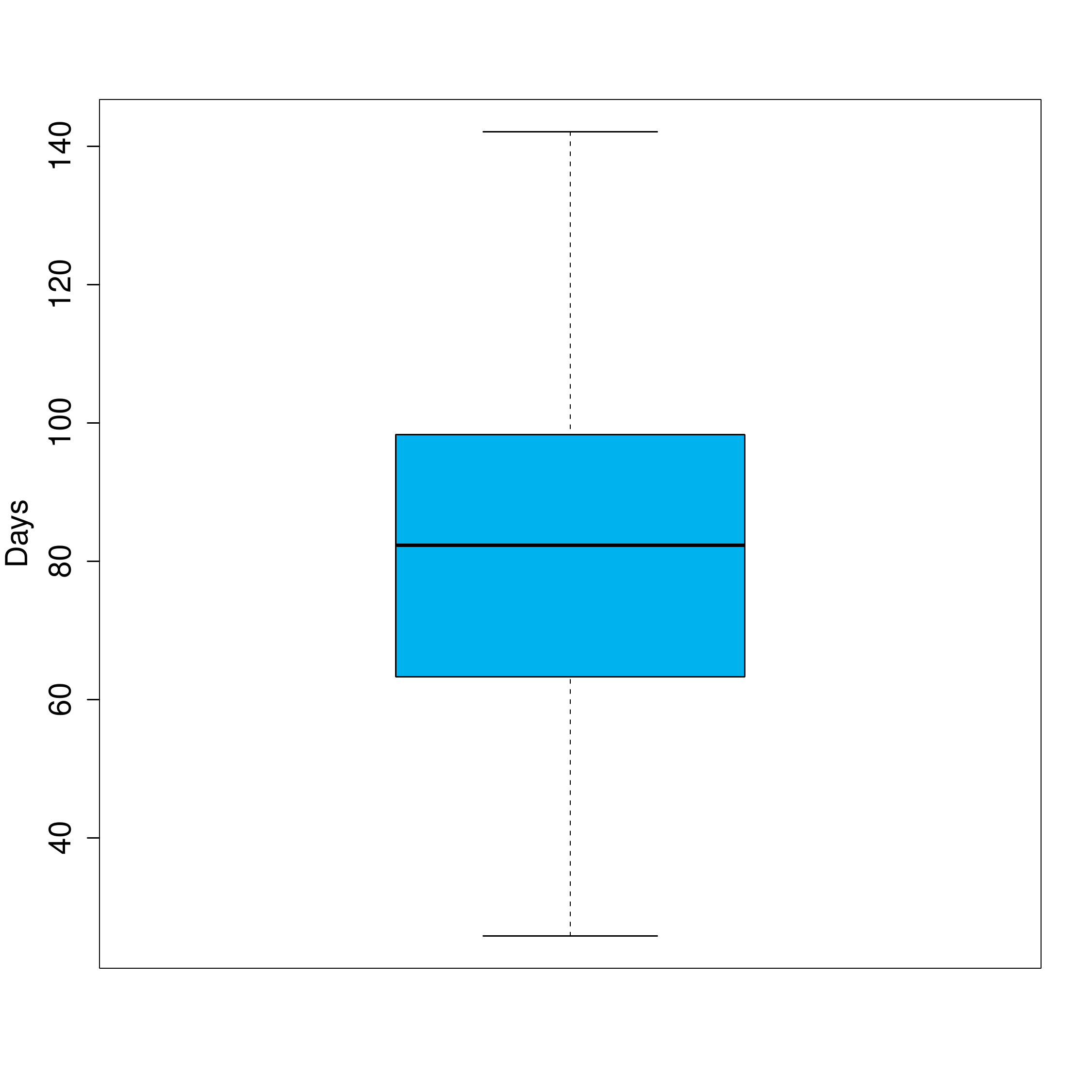}
	\caption{\small Left: Average daily temperatures for each of the $65$ weather stations. Right: Average number of sunny days per year for the $50$ weather stations with available records.}
	\label{fig:Sunny-Days-Data}
\end{figure}

\begin{figure}[!ht]
	\centering
	\includegraphics[width=.66\textwidth,clip,trim={0cm 0.5cm 0cm 1.25cm}]{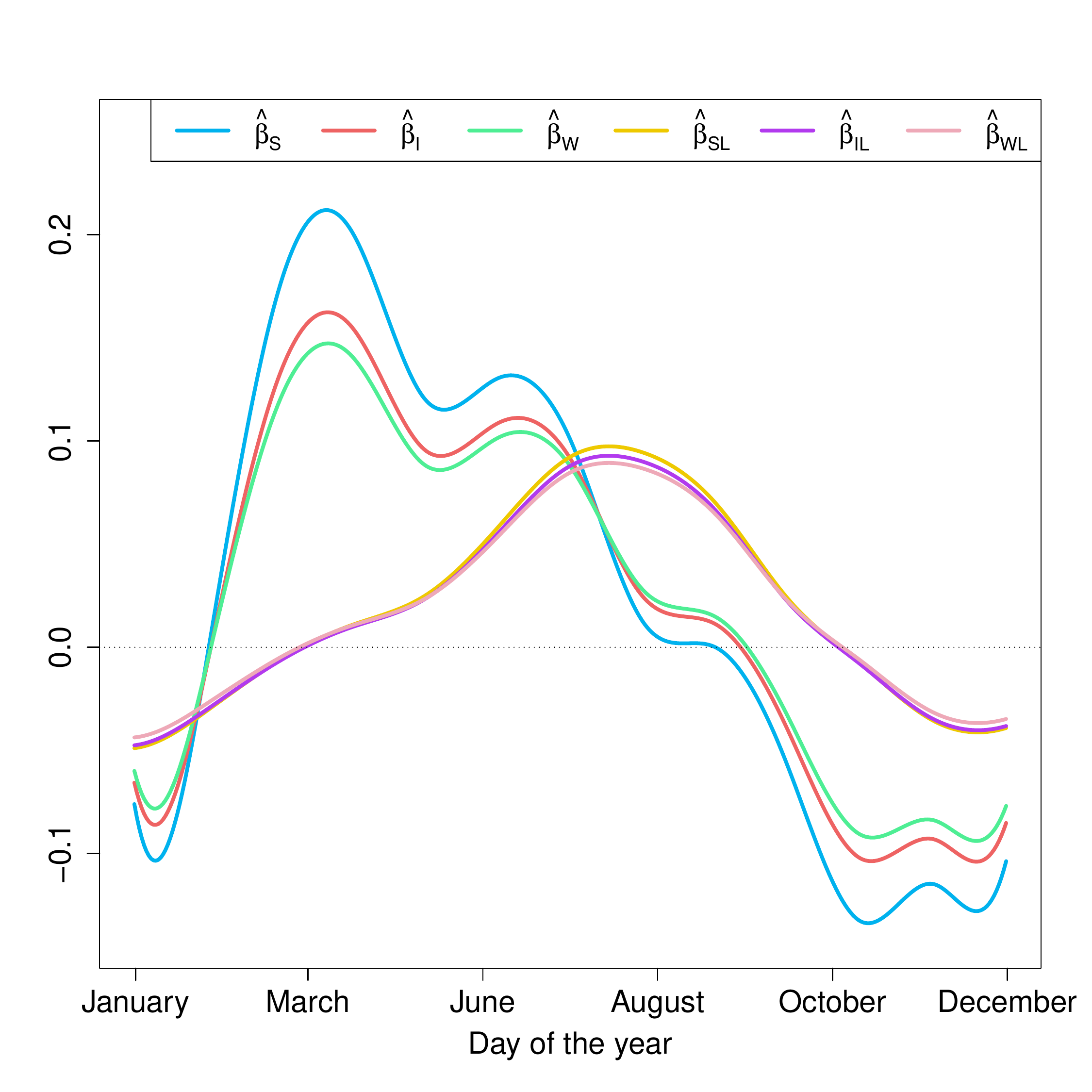}
	\caption{\small Estimators of the functional slope of the FLMSR.}
	\label{fig:Sunny-Days-Betas}
\end{figure}

The proposed testing procedure with the six considered estimators of the functional slope of the FLMSR is applied. As in the Monte Carlo experiments, $B=1000$ bootstrap samples are considered to compute the $p$-values associated with each estimator of $\beta$. The cutoff $K_{\max}=3$ is set since the first three FPCs of the average daily temperatures explain $99.63\%$ of the total variability of the temperatures. Then, the six estimators of $\beta$, shown in Figure~\ref{fig:Sunny-Days-Betas}, and their associated residuals are obtained. The difference between the OLS+CV-based estimators and the LASSO-based estimators is due to the fact that the former selects $3$ FPCs and the latter $2$ FPCs. Since fewer FPCs are used, the variability of the LASSO estimators is lower. Furthermore, it can be seen that the simplified OLS+CV estimator is somewhat different from the other two OLS+CV estimators, so the effect of the imputations seems to be relevant. Finally, Figure~\ref{fig:Sunny-Days-Statistics} shows the test statistics, the bootstrap statistics, and a kernel density estimator for the latter. In particular, the $p$-values for the estimators $\widehat{\beta}_{\mathcal{S}}$, $\widehat{\beta}_{\mathcal{SL}}$, $\widehat{\beta}_{\mathcal{I}}$, $\widehat{\beta}_{\mathcal{IL}}$, $\widehat{\beta}_{\mathcal{W}}$, and $\widehat{\beta}_{\mathcal{WL}}$ are, respectively, $0.386$, $0.248$, $0.369$, $0.243$, $0.359$, and $0.253$, indicating that the procedure does not reject the linearity hypothesis with any of the six estimators considered. Consequently, there is no evidence in the sample to reject FLMSR assumed in \cite{Febrero-Bande2019} to explain the average of the number of sunny days per year with the mean curve of the annual average daily temperature in the presence of missing responses.

\begin{figure}[!htpb]
	\centering
	\includegraphics[width=.46\textwidth,clip,trim={0cm 0.5cm 0cm 1.25cm}]{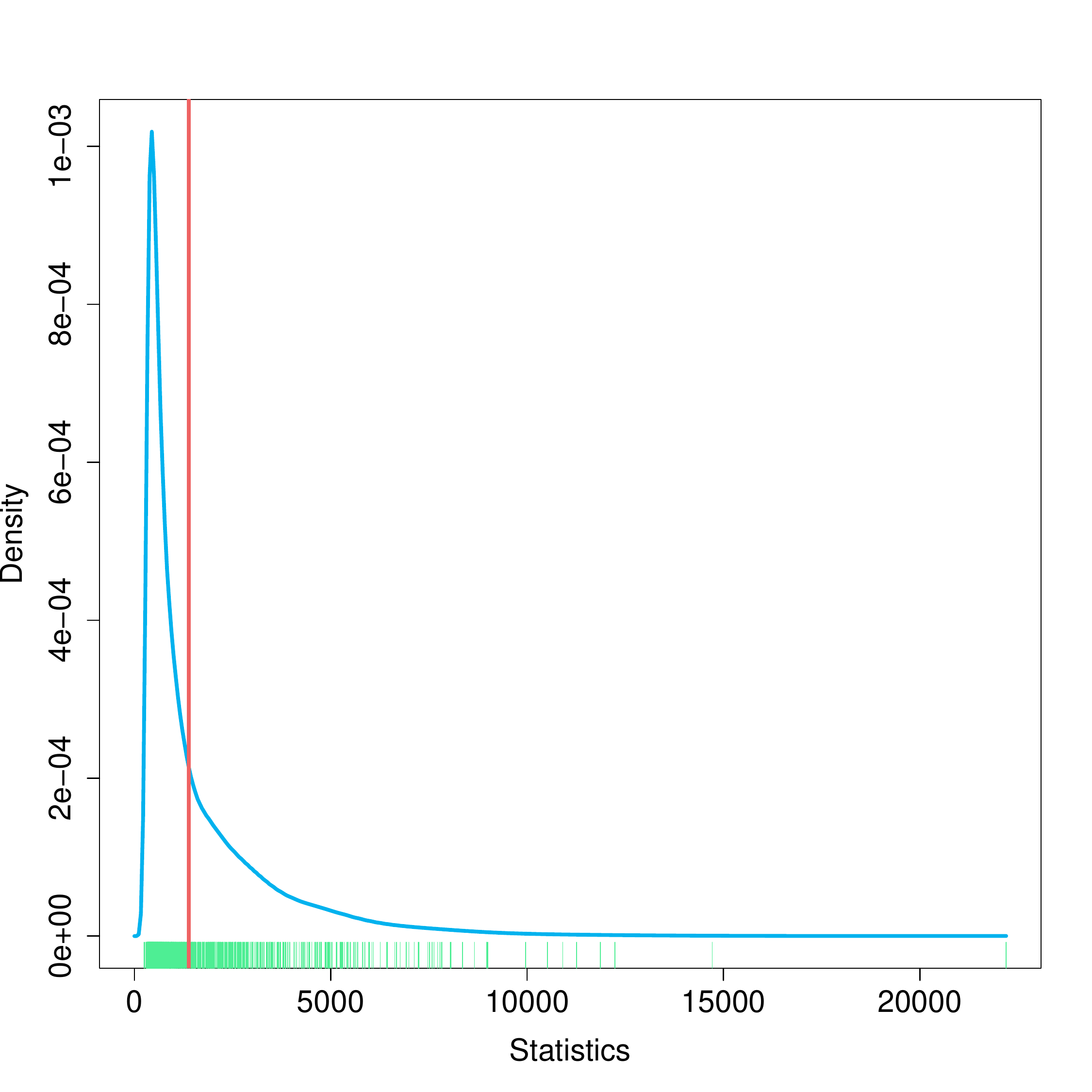}\includegraphics[width=.46\textwidth,clip,trim={0cm 0.5cm 0cm 1.25cm}]{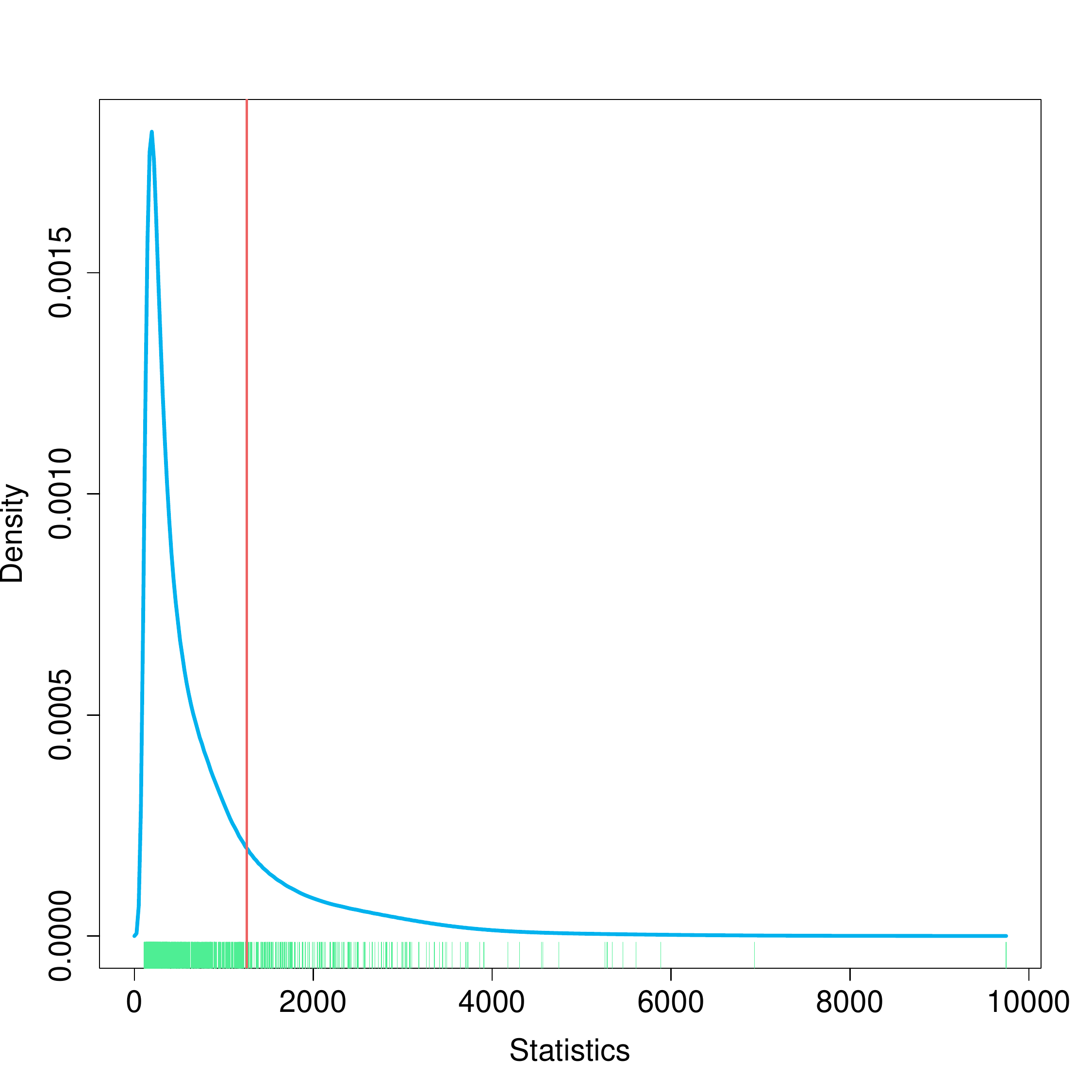}\\
	\includegraphics[width=.46\textwidth,clip,trim={0cm 0.5cm 0cm 1.25cm}]{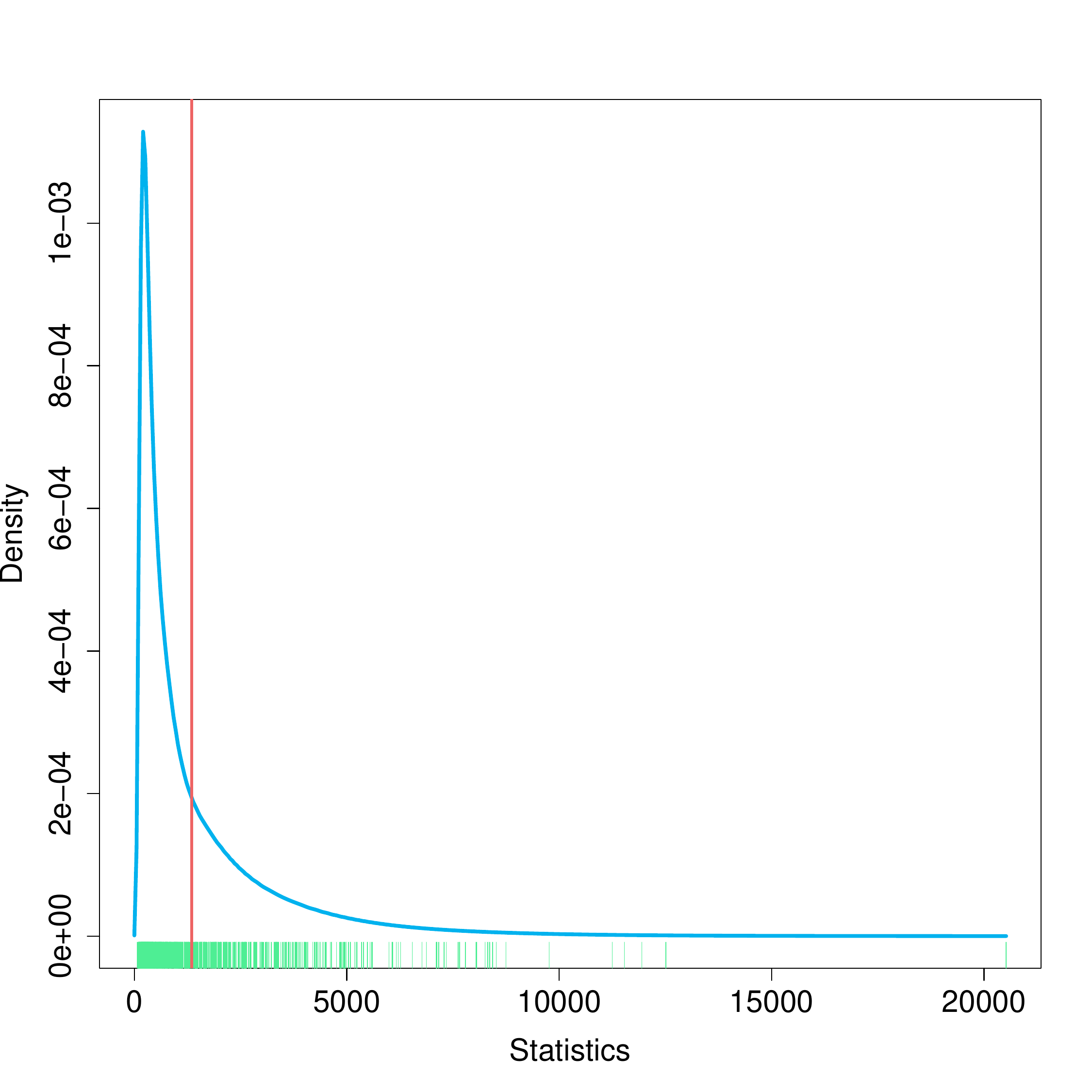}\includegraphics[width=.46\textwidth,clip,trim={0cm 0.5cm 0cm 1.25cm}]{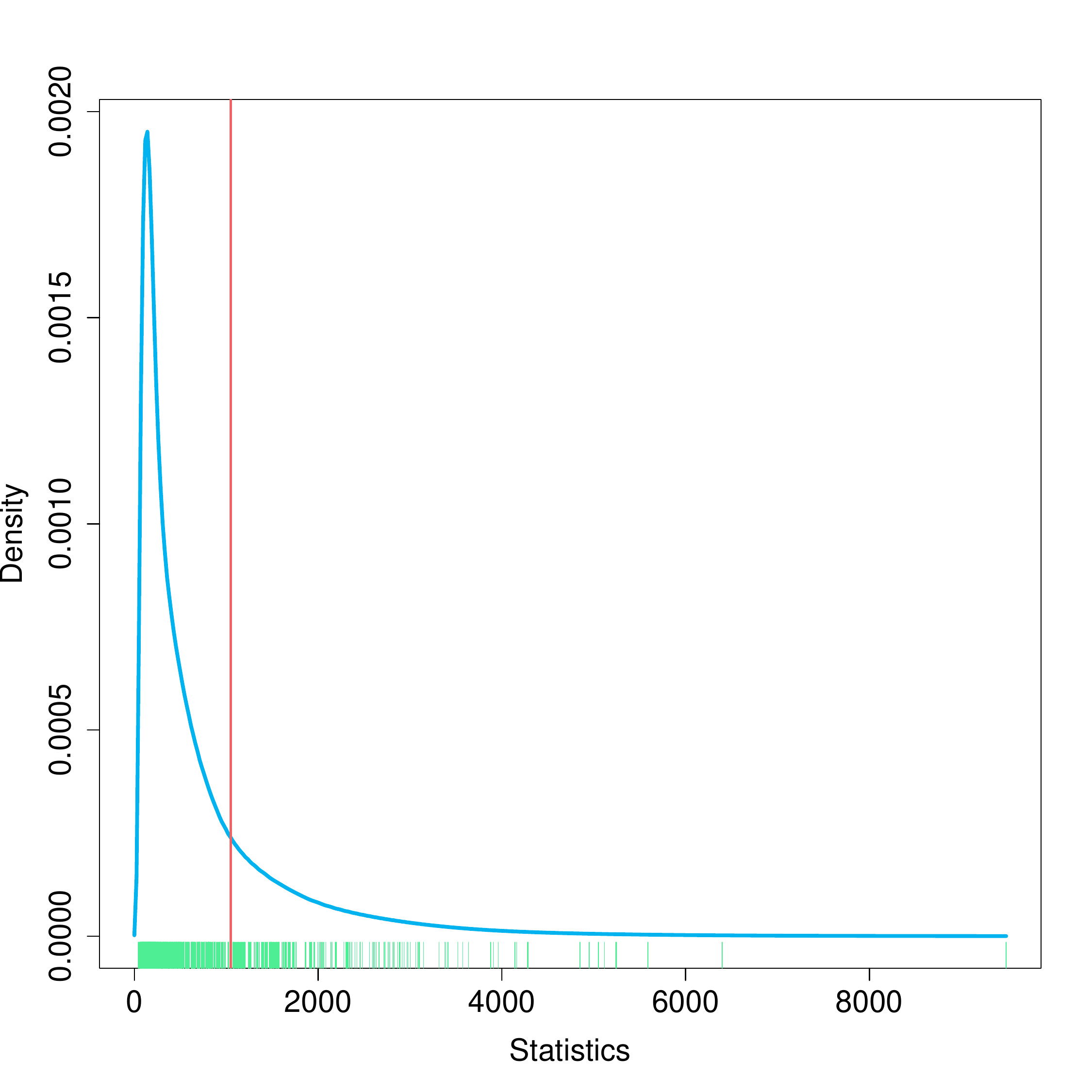}\\
	\includegraphics[width=.46\textwidth,clip,trim={0cm 0.5cm 0cm 1.25cm}]{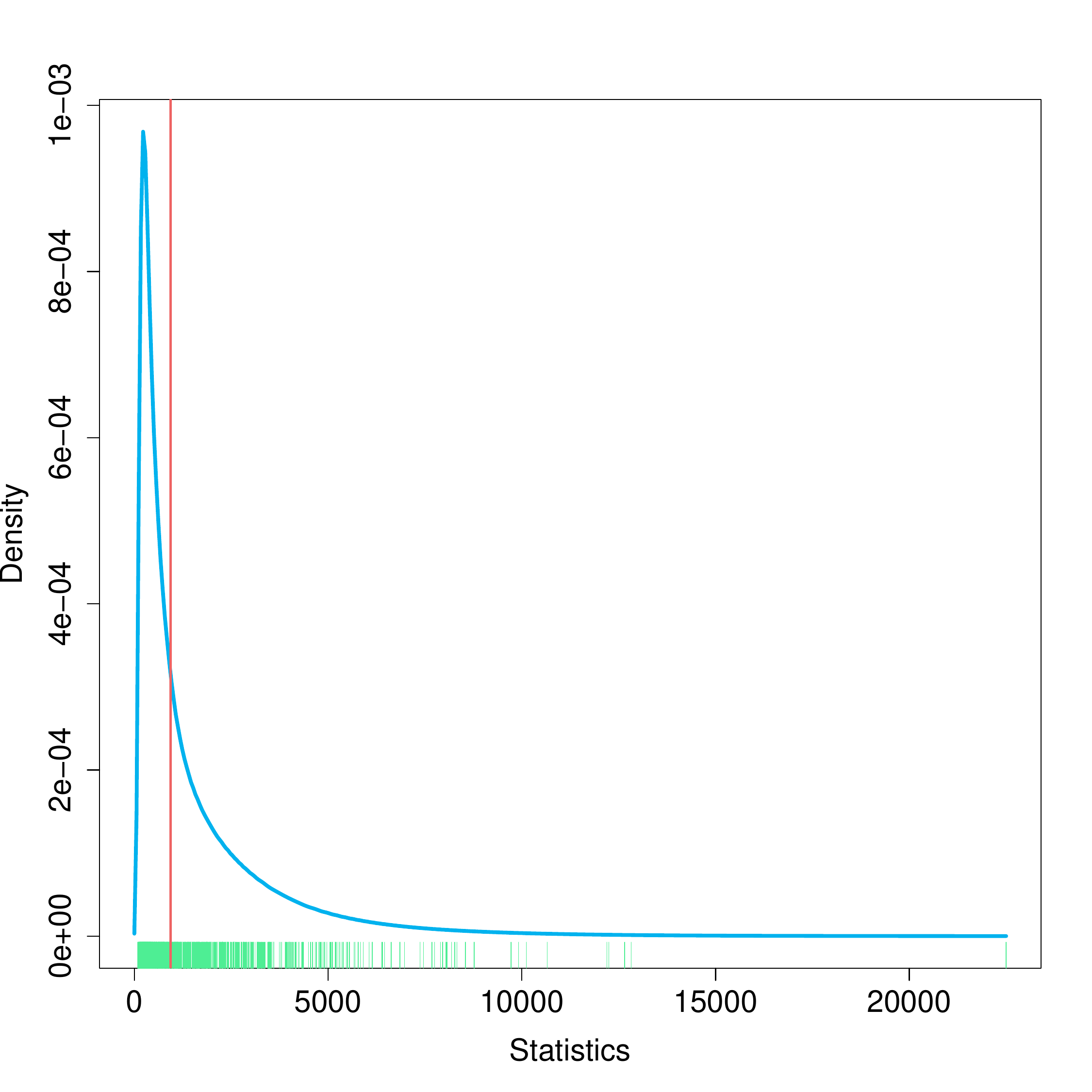}\includegraphics[width=.46\textwidth,clip,trim={0cm 0.5cm 0cm 1.25cm}]{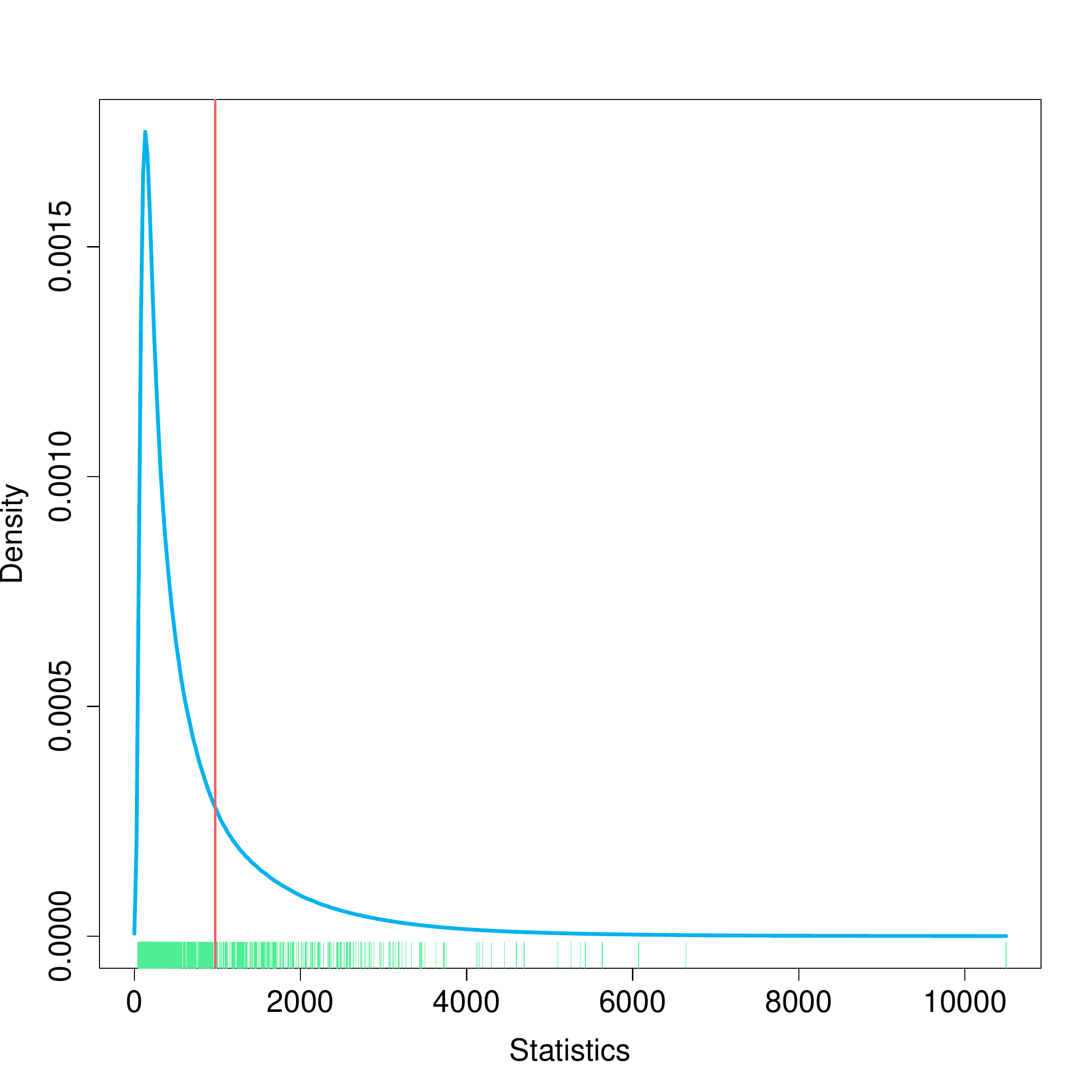}
	\caption{\small From left to right and from top to bottom, the panels show the test statistics (vertical lines) and the kernel density estimators of the bootstrapped statistics for the tests run with $\widehat{\beta}_{\mathcal{S}}$, $\widehat{\beta}_{\mathcal{SL}}$, $\widehat{\beta}_{\mathcal{I}}$, $\widehat{\beta}_{\mathcal{IL}}$, $\widehat{\beta}_{\mathcal{W}}$, and $\widehat{\beta}_{\mathcal{WL}}$, respectively. The respective $p$-values are $0.386$, $0.248$, $0.369$, $0.243$, $0.359$, and $0.253$.}\label{fig:Sunny-Days-Statistics}
\end{figure}

\section{Conclusions}
\label{sec:Conclusions}

A goodness-of-fit testing procedure for the null hypothesis of the FLMSR with MAR responses has been proposed. The procedure adapts the approach in \cite{Garcia-Portugues:flm} to consider MAR responses. Six different estimation methods for the functional slope of the model were considered, including those in \cite{Febrero-Bande2019}. The ones that seem to show the best performance in terms of testing in the Monte Carlo experiments are the imputed and inverse probability weighted estimators. Importantly, this evidences that testing after eliminating the pairs with missing responses is suboptimal. The Monte Carlo experiments also show that the MAR-adapted testing procedure behaves well in practice because it respects the significance level and has sufficient power to detect nonlinearities even when their effect is small. The procedure was applied to a real dataset to determine whether the FLMSR is sufficiently good for it, and no evidence was found to reject this hypothesis.

A limitation of the testing procedure is that it seems to be only useful for small and medium sample sizes; as the number of observations increases and the percentage of missingness is not high, the conducted Monte Carlo experiments show a small difference between testing on the complete pairs and testing also including pairs with missing responses. The computational cost of the procedure can be reduced by considering random projections as in \cite{Cuesta-Albertos:gofflm}, although a decrease in power is to be expected. As possible extensions of the proposed goodness-of-fit test, other regression models with functional covariates and scalar response are susceptible to be tackled with the same methodology if the corresponding MAR-estimation pipeline is developed. Other extensions may include the consideration of several functional predictors and/or a functional response within a functional linear model.

\section*{Acknowledgments}

The authors acknowledge financial support by MCIN/\-AEI/\-10.13039/\-501100011033: the first and fourth authors from grant PID2020-116587GB-I00, the second author from grant PID2019-108311GB-I00, and the third author from grant PID2021-124051NB-I00. Comments from two anonymous reviewers are acknowledged, as well as the assistance of the Associate Editor.


\end{document}

%% file: preamble_arXiv.tex
\usepackage[utf8]{inputenc}
\usepackage[T1]{fontenc}

\usepackage{amsthm}
\usepackage{amsmath}
\usepackage{amsfonts}
\usepackage{amssymb}

\usepackage{graphicx} 
\usepackage{float}
\usepackage{booktabs}
\usepackage{multirow}
\usepackage{rotating}
\usepackage{array}
\usepackage{xcolor}
\usepackage{subfig}
\usepackage[ruled]{algorithm2e}

\newcolumntype{R}[1]{>{\raggedleft\let\newline\\\arraybackslash\hspace{0pt}}m{#1}}
\newcolumntype{L}[1]{>{\raggedright\let\newline\\\arraybackslash\hspace{0pt}}m{#1}}

\usepackage{breakcites}

\usepackage{nowidow}

\usepackage{enumitem}

\usepackage[top=1.5cm,bottom=2cm,right=2.5cm,left=2.5cm]{geometry}
\usepackage{titling}

\usepackage{natbib}

\usepackage{ifplatform}

\usepackage{textcomp}
\usepackage{bbm}

\ifwindows
\fi

\allowdisplaybreaks

\newcommand{\Xcal}{\mathcal{X}}
\newcommand{\xcal}{{\scriptstyle\mathcal{X}}}
\DeclareFontFamily{OT1}{pzc}{}
\DeclareFontShape{OT1}{pzc}{m}{it}{<-> s * [1.10] pzcmi7t}{}
\DeclareMathAlphabet{\mathpzc}{OT1}{pzc}{m}{it}

\newtheorem{lemma}{Lemma}
\newtheorem{algo}{Algorithm}

